\documentclass[12pt]{IEEEtran}
\usepackage{graphicx, amssymb}
\usepackage{graphicx}
\newtheorem{thm}{Theorem}[section]
\newtheorem{lem}[thm]{Lemma}
\newtheorem{cor}[thm]{Corollary}
\newtheorem{prop}[thm]{Proposition}

\newtheorem{assum}[thm]{Assumption}

\makeatletter
\def\ifundefined{\@ifundefined}
\makeatother

\begin{document}

\title{Performance Analysis of Iterative Channel Estimation and Multiuser Detection in Multipath DS-CDMA Channels}

\author{Husheng Li, Sharon M. Betz and H. Vincent Poor\thanks{Husheng Li is
with Qualcomm Inc., San Diego, CA, 92121, USA (email:
hushengl@qualcomm.com).  Sharon M. Betz is with Department of
Electrical Engineering, Princeton University, Princeton, NJ 08544,
USA (email: sbetz@princeton.edu). H. Vincent Poor is with
Department of Electrical Engineering, Princeton University,
Princeton, NJ 08544, USA (email: poor@princeton.edu). This
research was supported by the Air Force Research Laboratory under
Cooperative Agreement No. FA8750-05-2-01 92.}}

% The major version number of the class file will not
% be defined with the old IEEEtran.cls. So, we can use this fact
% to determine if we are running the old or the new class.
\ifundefined{IEEEtransversionmajor}{%
   % This block will be executed only if we are using the old
   % class. All we do is to make sure the V1.3 lengths and commands
   % actually exist so the code won't choke when it
   % doesn't find them.

   % This file doesn't need most of these definitions.
   % However, we'll provide them all in case somebody
   % wants to see what should be executed when compiling
   % a V1.3 or later IEEEtran.cls .tex file with a pre V1.3
   % IEEEtran.cls class file. In such a case, all you have to
   % do is copy this block to the start of your code.
   % However, it won't fix any bugs in the old IEEEtran.cls!
   %
   % **** BACKWARD COMPATIBILITY CODE BLOCK START ****
   \newlength{\IEEEilabelindent}
   \newlength{\IEEEilabelindentA}
   \newlength{\IEEEilabelindentB}
   \newlength{\IEEEelabelindent}
   \newlength{\IEEEdlabelindent}
   \newlength{\labelindent}
   \newlength{\IEEEiednormlabelsep}
   \newlength{\IEEEiedmathlabelsep}
   \newlength{\IEEEiedtopsep}

   \providecommand{\IEEElabelindentfactori}{1.0}
   \providecommand{\IEEElabelindentfactorii}{0.75}
   \providecommand{\IEEElabelindentfactoriii}{0.0}
   \providecommand{\IEEElabelindentfactoriv}{0.0}
   \providecommand{\IEEElabelindentfactorv}{0.0}
   \providecommand{\IEEElabelindentfactorvi}{0.0}
   \providecommand{\labelindentfactor}{1.0}

   \providecommand{\iedlistdecl}{\relax}
   \providecommand{\calcleftmargin}[1]{
                   \setlength{\leftmargin}{#1}
                   \addtolength{\leftmargin}{\labelwidth}
                   \addtolength{\leftmargin}{\labelsep}}
   \providecommand{\setlabelwidth}[1]{
                   \settowidth{\labelwidth}{#1}}
   \providecommand{\usemathlabelsep}{\relax}
   \providecommand{\iedlabeljustifyl}{\relax}
   \providecommand{\iedlabeljustifyc}{\relax}
   \providecommand{\iedlabeljustifyr}{\relax}

   \newif\ifnocalcleftmargin
   \nocalcleftmarginfalse

   \newif\ifnolabelindentfactor
   \nolabelindentfactorfalse

   % in V1.4 of IEEEtran.cls
   \newif\ifcenterfigcaptions
   \centerfigcaptionsfalse

   % we need to provide the old IED environments
   % with a bogus optional argument
   \let\OLDitemize\itemize
   \let\OLDenumerate\enumerate
   \let\OLDdescription\description

   \renewcommand{\itemize}[1][\relax]{\OLDitemize}
   \renewcommand{\enumerate}[1][\relax]{\OLDenumerate}
   \renewcommand{\description}[1][\relax]{\OLDdescription}

   \providecommand{\pubid}[1]{\relax}
   \providecommand{\pubidadjcol}{\relax}
   \providecommand{\specialpapernotice}[1]{\relax}
   \providecommand{\overrideIEEEmargins}{\relax}

   % V1.1 change: use \let instead of \providecommand
   % This prevents LaTeX from hanging if the user ever
   % tried to redefine \PARstart in terms of \CMPARstart
   \let\CMPARstart\PARstart

   \let\OLDappendix\appendix
   \renewcommand{\appendix}[1][\relax]{\OLDappendix}

   \newif\ifuseRomanappendices
   \useRomanappendicestrue

   % V1.2 change: handle the optional biography environment argument
   % (the photo specifier) provided by IEEEtran V1.5 and later.
   % This is tricky because, under the new biography, the SECOND
   % argument is the non-optional one (the biography text).
   \let\OLDbiography\biography
   \let\OLDendbiography\endbiography
   \renewcommand{\biography}[2][\relax]{\OLDbiography{#2}}
   \renewcommand{\endbiography}{\OLDendbiography}
   % **** BACKWARD COMPATIBILITY CODE BLOCK END ****
   % alter the header to show we are using the older class
   \markboth{A Test for IEEEtran.cls--- {\tiny \bfseries
   [Running Older Class]}}{Shell: A Test for IEEEtran.cls}}{
   % END IF OLDER CLASS

   % This block will be executed only if we are running
   % the enhanced class
   % alter the header to show we are using the enhanced class
   \markboth{}%
   {Shell: A Test for IEEEtran.cls}}

% end of conditional

% Uncomment this line to render the big first letter in
% Computer Modern font.
%\renewcommand{\PARstart}[2]{\CMPARstart{#1}{#2}}
% V1.1 change: This would hang if using the older IEEEtran.cls
% in IEEEtest V1.0, it is OK now.
%
%
% Here's how you would do invited papers:
%\specialpapernotice{(Invited Paper)}
% If you are binding copies of work generated with IEEEtran.cls,
% you may want to try:
%\overrideIEEEmargins
% These commands work OK here even though they are not in the preamble.
%(I wanted to put them after the backward compatibility code)

\maketitle
% here's how you get a publisher's ID mark with the new
% IEEEtran.cls.  If you want to use it, don't forget to
% also uncomment the \pubidadjcol command (which must be
% executed in the second text column) around line 434 below
%\pubid{0000--0000/00\$00.00~\copyright~2001 IEEE}

\begin{abstract}
This paper examines the performance of decision
feedback based iterative channel estimation and multiuser
detection in channel coded aperiodic DS-CDMA systems operating over multipath fading
channels. First, explicit expressions describing the performance of channel
estimation and parallel interference cancellation based multiuser detection are developed. These
results are then combined to characterize
the evolution of the performance of a system that iterates among channel
estimation, multiuser detection and channel
decoding. Sufficient conditions for convergence of this system to a unique
fixed point are developed.
\end{abstract}

% no need for this single page document
% you may have to move \pubidadjcol (if used) if
% these are enabled
%\listoffigures
%\listoftables
%\tableofcontents
\section{Introduction}
Direct sequence code division multiple-access (DS-CDMA) has been
selected as the fundamental signaling technique for third generation
(3G) wireless communication systems, due to its advantages of soft
user capacity limit and inherent frequency diversity. However, it
suffers from multiple-access interference (MAI) caused by the
non-orthogonality of spreading codes, particularly for heavily
loaded systems. Therefore, techniques for mitigating the MAI, namely
multiuser detection, have been the subject of an intensive research
effort over the past two decades. It is well known that multiuser
detection can substantially suppress MAI, thus improving system
performance. Maximum likelihood (ML) %based
 multiuser
detection~\cite{Verdu1986} was proposed in the early 1980s, and
achieves the optimal performance at the cost of prohibitive
computational cost when the number of users is large. For practical
implementation, suboptimal algorithms, such as the linear minimum
mean square error (LMMSE) detector \cite{Lupas1989} or decorrelator
\cite{Verdu1998}, allow a tradeoff between complexity and
performance. It should be noted that the technique of multiuser
detection is being applied in existing CDMA systems, such as EV-DO
Revision A systems \cite{Hou2006}.

In recent years, the turbo principle, namely the iterative exchange
of soft information among different blocks in a communication system
to improve the system performance, has been applied to combine
multiuser detection with channel decoding
\cite{Alex2000}\cite{Moher}\cite{Poor2004}\cite{Reed1998}\cite{Valenti1998}\cite{wang1999}.
In such turbo multiuser detectors, the outputs of channel decoders
are fed back to the multiuser detector, thus enhancing the
performance iteratively. Turbo multiuser detection based on the
maximum \textit{a posteriori} probability (MAP) detection and
decoding criterion has been proposed in \cite{wang2004}\cite{wang1999} together
with a lower complexity technique based on interference cancellation
and LMMSE filtering. Further simplification is obtained by applying
parallel interference cancellation (PIC)~\cite{Alex2000} for
multiuser detection, where the decisions of the decoders are
directly subtracted from the original signal to cancel the MAI.

Practical wireless communication systems usually experience fading
channels, whose state information is unknown to the receiver. Thus
practical systems need to consider detection and decoding with
uncertain channel state information. In the context of short code
CDMA systems, blind multiuser detection can be accomplished without
explicit channel estimation by using subspace and other techniques
\cite{wang2000}. An alternative receiver structure adopts an
explicit channel estimation block and carries out the decoding with
the corresponding channel estimate. In systems without decision
feedback, the channel estimation block is cascaded with the decoder
and operates as a front end for the subsequent blocks. With such a
receiver structure, the channel estimates can be obtained with
training symbols~\cite{Buzzi2001} or with blind estimation
algorithms~\cite{Xu2000}. Explicit expressions for the performance
of such channel estimation schemes are given in~\cite{LiHuCISS2003}
and the corresponding impact on multiuser detection is discussed in
the large system limit in \cite{Evans2000}
and~\cite{LiHuEuroSip2005}. In systems with decision feedback, the
decisions of the decoder are fed back to the channel estimator to
enhance its performance. In such systems, the channel estimator and
the decoder can operate either simultaneously~\cite{Raheli1995} or
successively~\cite{Buzzi2003}~\cite{Kobayashi2001}~\cite{Niu2003}.
An example of the former strategy applied to ML sequence detection
in uncertain environments is proposed in~\cite{Raheli1995}; called
per-survivor processing, tentative decisions are immediately fed
back to the channel estimation algorithm and the corresponding
estimates are used for the detection of future symbols.
%As an example of the former strategy, per-survivor processing,
%where tentative decisions are immediately fed back to the channel
%estimation algorithm and the corresponding estimates are used for
%the detection of future symbols, is proposed in~\cite{Raheli1995}
%for ML sequence detection in uncertain environments.
In the latter strategy, the decisions are fed back only when the
entire current decoding procedure is finished. For example,
in~\cite{Kobayashi2001}, an expectation maximization (EM) %based
channel estimation algorithm, combined with successive
interference cancellation, is proposed. Joint channel estimation
and data detection algorithms for uncoded single-antenna and
multiple-antenna systems are discussed in~\cite{Buzzi2004}
and~\cite{Buzzi2003}, respectively. In channel coded systems,
iteration can achieve better performance when the turbo principle
is applied, due to the redundancy introduced by the code
structure. In~\cite{Niu2003}, an iterative algorithm is proposed
and analyzed for channel estimation and decoding of low-density
parity-check (LDPC) coded quadrature amplitude modulation (QAM)
systems.

In this paper, we consider channel-coded CDMA systems operating
over multipath fading channels whose channel state information is
unknown to the receiver. To demodulate and decode such systems, we
apply the turbo principle to both channel estimation and multiuser
detection. As shown in Figure 1, we consider a receiver that feeds
back decisions from channel decoders to both an ML %based
 channel
estimator and a PIC %based
 multiuser detector. The iteration is
initialized with training symbol based channel estimation and a
non-iterative multiuser detection. The receiver structure is
similar to those proposed in
\cite{AlexISSS2000}\cite{Lampe2002}\cite{Loncar2004}. However,
this paper is focused mainly on the performance analysis of such
structures using semi-analytic methods. We analyze the
contributions to the variance of the channel estimation error due
to noise and decision feedback error, and the variance of the
residual MAI after PIC. We then use this analysis to describe the
decoding process as an iterative mapping. We also propose
conditions assuring convergence of this iterative mapping to a
unique fixed point. We further compute the asymptotic multiuser
efficiency (AME)~\cite{Verdu1998} of this overall system, under
some mild assumptions on the channel decoders. It should be noted
that the analysis in this paper is based on large sample and large
system analysis.

The remainder of this paper is organized as follows. Section II
introduces the signal model and the channel decoder used in our
analysis. The performance analyses of ML channel estimation and PIC
multiuser detection are given in Section III and Section IV,
respectively. Based on these results, the corresponding iterative
mapping is described and analyzed in Section V. Numerical results
and conclusions are given in Section VI and Section VII,
respectively. The notations used in this paper are explained as
follows.
\begin{itemize}
\item Throughout this paper, if no special note is given, we denote
vectors with small letters in bold fonts, matrices with capital
letters in bold fonts and scalars with non-bold fonts. 

\item For any
variable $X$, we denote the corresponding estimate from the decision
feedback by $\hat{X}$ and the corresponding error $X-\hat{X}$ by
$\delta X$.

\item Superscript $T$ denotes transposition and superscript $H$
denotes conjugate transposition.

\item  $\mathbf{I}$ denotes the identity
matrix.

\item $\lceil x\rceil$ denotes the smallest
integer larger than or equal to $x$. 

\item mod$(i,j)$ denotes the modulo
of $i$ with respect to $j$, with the convention of mod$(i,i)=i$.

\item For a matrix $\mathbf{A}_{m\times n}$, $\left\|\mathbf{A}\right\|_F\triangleq
\sqrt{\sum_{i=1}^m\sum_{j=1}^nA_{ij}^2}$ is the Frobenius norm of
$\mathbf{A}$.
\end{itemize}

\section{Signal Model}
\subsection{Signal Model}
We consider a synchronous uplink long code (aperiodic) DS-CDMA
system, with identical channel coding, binary phase-shift keying
(BPSK) modulation, $K$ active users, spreading gain $N$, system load
$\beta=\frac{K}{N}$, and identical transmission rates for all users.
The transmitted symbols experience multipath fading. We adopt a
block fading model and denote by $M$ the coherence time, measured in
the number of symbol periods, over which the channel is stationary.
Within a coherence period, the chip matched filter output of the
receiver at symbol period $t$ can be collected into an $N$-vector
given by
\begin{eqnarray}\label{signalmodel}
 \mathbf{r}(t) = \sum_{k=1}^K b_k(t)
 \sum_{l=1}^La_{kl}\mathbf{s}_{kl}(t)+\mathbf{n}(t),\qquad
 t=1,2,...M,
\end{eqnarray}
where $L$ denotes the number of resolvable paths per user, $b_k(t)$
denotes the channel coded binary symbols, $a_{kl}$ denotes the
channel gain of the $l$-th path of user $k$, $\mathbf{s}_{kl}(t)$
denotes the binary spreading code with
$\left\|\mathbf{s}_{kl}(t)\right\|=1$ received from user $k$ along
path $l$ at time $t$ and $\mathbf{n}(t)$ is an $N$-vector of
independent and identical distributed (i.i.d.)
 circularly symmetric complex Gaussian (CSCG)
\footnote{A complex random variable is CSCG distributed if its real
and imaginary parts are mutually independent Gaussian random
variables with zero mean and identical variance.}noise variables
with (normalized) variance $\sigma_n^2$. It should be noted that
although the assumption of synchronicity is valid in time division
duplexing (TDD) systems, it does not hold for many frequency
division duplexing (FDD) systems. However, as it will be shown, the results from the
analysis of synchronous systems are also reasonably valid, though not exactly the same, in the case of
asynchronous systems.

For the system model, we have the following assumptions.
\begin{assum}\label{Assum1}
The channel gains $\left\{a_{kl}\right\}$ are independently CSCG
distributed with zero means and variances $\frac{1}{L}$. We consider
only the case of large $L$, which implies that
$\sum_{l=1}^L|a_{kl}|^2\approx 1$, $k=1,...,K$; thus all users
achieve the same performance with maximal ratio combining (MRC).
\end{assum}
\begin{assum}\label{Assum2}
We ignore intersymbol interference (ISI) and assume that the
spreading codes received along different paths of a given user are
mutually independent (\textit{independent model}).
\end{assum}
\begin{assum}\label{Assum3}
Based on Assumption {\ref{Assum2}}, the crosscorrelations
$\rho_{klmn}(t)\triangleq \mathbf{s}_{kl}(t)^T\mathbf{s}_{mn}(t)$
(note that $\rho_{klkl}(t)=1$) satisfy
\begin{itemize}
\item  $E\left\{\rho_{klmn}(t)\right\}=0$, if $(k,l)\neq
(m,n)$;
\item  $E\left\{\rho_{klmn}^2(t)\right\}=\frac{1}{N}$, if $(k,l)\neq
(m,n)$;
\item  $E\left\{\rho_{klmn}(t)\rho_{pqrs}(t)\right\}=0$,
if $(k,l,m,n)\neq (p,q,r,s)$.
\end{itemize}
\end{assum}

The above assumptions simplify the performance analysis
substantially. Moreover, these assumptions are reasonable for
practical systems due to the following reasons:
\begin{itemize}
\item Assumption \ref{Assum1} is based on the fact that more propagation paths
are resolvable in CDMA systems than narrow band systems,
particularly in environments with abundant scattering (e.g., indoor
environment). With this assumption, we ignore the impact of the
fluctuation of received power incurred by the multipath fading, and
consider only the impairment caused by the channel estimation error.

\item Assumption \ref{Assum2} is unrealistic since these sequences
are shifted versions of each other (\textit{shifted model}).
However, the accuracy of the results dependent upon this
assumption is validated with numerical results in Section VI and
asymptotic analysis given in Appendix 1.
\end{itemize}

\subsection{Receiver Structure}
The structure of receiver is shown in Figure 1. The channel
coefficients are estimated in the channel estimator, which operates in a
`semi-blind' way. Training symbols are available to obtain an
initial estimate in the first iteration. In the further iterations
the information symbol decisions from channel decoders are assumed
to be correct. Then, both the training symbols and fed back decisions are considered as training symbols and used for ML channel
estimation. A multiuser detector is used to mitigate the MAI and its
outputs are de-interleaved and decoded in the channel decoder. In
the multiuser detector, we use the LMMSE algorithm in the first
iteration and the PIC algorithm with the aid of hard decision
feedback in the succeeding iterations. We follow the standard
procedure in turbo multiuser detection
\cite{Alex2000}\cite{Kobayashi2001}\cite{Moher}\cite{wang2004} to
reconstruct the channel symbols from the channel decoder output.
Then these channel symbol estimates are interleaved and fed back to
the multiuser detector and channel estimator to enhance the
performance iteratively.

We denote by $\hat{b}_k(t)$ the estimated binary channel symbol of
user $k$ at symbol period $t$ that is fed back from the channel
decoder. For simplicity, we use hard decision feedback and denote
the feedback symbol error rate by $P_e$. The decision feedback
error is denoted by $\delta b_k(t)\triangleq b_k(t)-\hat{b}_k(t)$.
Supposing that both $b_k(t)$ and $\delta b_k(t)$ are symmetrically
distributed, it is easy to check that
\begin{itemize}
\item $E\left\{\delta b_k(t)\right\}=0$;
\item $E\left\{b_k(t)\delta b_k(t)\right\}=2P_e$;
\item $E\left\{\delta b^2_k(t)\right\}=4P_e$.
\item $E\left\{\delta b_k(m)\delta b_l(n)\right\}=0$, when $(k,m)\neq (l,n)$.
\end{itemize}

It should be noted that, in practical systems, soft decision
feedback will achieve better performance than hard decision
feedback. However, the performance of channel estimation with soft
decision feedback is determined by both the first and second moments
of the decision feedback error~\cite{LiHuCISS2003}. Thus the
corresponding analysis of performance evolution is more complicated
than the case of hard decision feedback. Therefore, we adopt
hard decision feedback in order to simplify the system performance
analysis.

For the decision feedback from channel decoders, we have the
following reasonable assumption, which simplifies the analysis and
is also used in \cite{Alex2000}.
\begin{assum}\label{Assum4}
The codeword length is assumed to be large enough so that the transmitted
symbols are coded over many coherence periods. The decision
feedbacks $\left\{\hat{b}_k(t)\right\}$ are mutually independent for
different $k$ or $t$.
\end{assum}

\section{Performance Analysis of Channel Estimation}
In this section, we discuss the performance of channel estimation.
First, we explain the training symbol based ML channel estimation
algorithm that is used in the first iteration. Then, we consider
the estimation of the channel coefficients with only hard decision
feedback from the channel decoders. %The performance results will be
%extended to channel estimation with both training symbols and
%decision feedback, which is used in the further iterations.
Finally, we extend the performance results to channel estimation
with both training symbols and decision feedback, the latter of
which is used in the further iterations.

In applying the turbo principle, to avoid the reuse of
information, only observations $\{\mathbf{r}(t)\}_{t\neq i}$ are
used in the channel estimation for multiuser detection in symbol
period $i$. Thus the corresponding channel estimation error is
independent of $\mathbf{r}(i)$. However, for simplicity of
discussion, we still assume that all $M$ received signals are used
for the channel estimation while retaining this independence
assumption. For large $M$, this results in only a small error in
the analysis.

In the following discussion of channel estimation and PIC, we
regard the channel gains $\{a_{kl}\}$ and the spreading codes
$\{\mathbf{s}_{kl}\}$ as realizations of random variables. Only
the transmitted symbols, decision feedback errors and noise are
considered as random variables. Throughout this paper, all
expectations, denoted as $E\{\cdot\}$, are over the distributions
of these three variables. Thus our results are conditioned on the
realizations of $\{a_{kl}\}$ and $\{\mathbf{s}_{kl}\}$. However,
by the strong law of large numbers, we will see that we can obtain
identical results for almost every realization of $\{a_{kl}\}$ and
$\{\mathbf{s}_{kl}\}$ in the large system limit
($K,N\rightarrow\infty$).

\subsection{Training Symbol Based ML Channel Estimation}
%We first assume that the $M$ channel symbols within a coherence
%period are all training symbols, which are known to the receiver.
%For simplicity, we incorporate the transmitted training symbols
%into the spreading codes and stack the received signals within
%this coherence period; thus (\ref{signalmodel}) can be rewritten
%as
First we assume that there are $M$ training symbols, channel
symbols known to the receiver, within a single coherence period.
For simplicity in deriving the channel estimate, we stack the chip
matched filter output of the signal corresponding to these
training symbols, rewriting (\ref{signalmodel}) as
\begin{eqnarray}
   \mathbf{r}=\mathbf{S}\mathbf{a}+\mathbf{n},
\end{eqnarray}
where
\begin{eqnarray}
   &\mathbf{r}&=\left(\mathbf{r}^H(1),..., \mathbf{r}^H(M)\right)^H_{NM\times 1},   \nonumber \\
   &\mathbf{n}&=\left(\mathbf{n}^H(1),..., \mathbf{n}^H(M)\right)^H_{NM\times 1},   \nonumber \\
   &\mathbf{a}&=\left(a_{11},a_{12},...,a_{KL}\right)_{KL\times 1}^T,   \nonumber \\
   &\mathbf{S}&=\left(\left(\mathbf{S}(1)\mathbf{B}(1)\right)^T,...,\left(\mathbf{S}(M)\mathbf{B}(M)\right)^T\right)_{NM\times KL}^T,
   \nonumber\\
   &\mathbf{B}(m)&=\left(
                     \begin{array}{cccc}
                       b_1(m)\mathbf{I}_{L\times L} & 0 & \cdots & 0 \\
                       0 & b_2(m)\mathbf{I}_{L\times L} & \cdots & 0 \\
                       \vdots & \vdots & \ddots & \vdots \\
                       0 & 0 & \cdots & b_K(m)\mathbf{I}_{L\times L} \\
                     \end{array}
                   \right)_{KL\times KL}\nonumber\\
   &\mathbf{S}(m)&=\left(\mathbf{s}_{11}(m),\mathbf{s}_{12}(m),...,\mathbf{s}_{KL}(m)\right)_{N\times KL},\qquad
   m=1,...,M.\nonumber
\end{eqnarray}

Applying the ML criterion and the normality of the noise, we can obtain
the ML channel estimate, which is given by
\begin{eqnarray}\label{MLestimation}
   \hat{\mathbf{a}}&=&\arg\max_{\mathbf{a}} P(\mathbf{r}|\mathbf{a})  \nonumber  \\
               &=&\arg\min_{\mathbf{a}}
               \|\mathbf{r}-\mathbf{S}\mathbf{a}\|\nonumber\\
               &=&(\mathbf{S}^T\mathbf{S})^{-1}\mathbf{S}^T\mathbf{r}  \nonumber  \\
               &=&\mathbf{R}^{-1}\mathbf{y},
\end{eqnarray}
where $\mathbf{R}=\mathbf{S}^T\mathbf{S}$ and
$\mathbf{y}=\mathbf{S}^T\mathbf{r}$.

It follows directly that the channel estimation error is
\begin{eqnarray}
   \delta\mathbf{a}&=&\mathbf{a}-\hat{\mathbf{a}} \nonumber\\
   &=&-\mathbf{R}^{-1}\mathbf{S}^T\mathbf{n},\nonumber
\end{eqnarray}
from which it is obvious that this error has zero mean and covariance
${\bf{\Sigma}_a}\triangleq E\left\{\delta \mathbf{a}\delta
\mathbf{a}^H\right\} =\sigma_n^2{\mathbf{R}}^{-1}$.

For a finite $M$, we can compute
$\mbox{trace}\left\{{\mathbf{R}}^{-1}\right\}$ in the large system
limit (i.e. when $K,N\rightarrow\infty$ while keeping
the system load, $\frac{K}{N}=\beta$, constant). For a system with system load $\beta$, it is
well known that as $K\rightarrow \infty$,
$\frac{K}{\mbox{trace}\left\{\hat{\mathbf{R}}^{-1}\right\}}$
converges to the multiuser efficiency of a decorrelator, namely
$1-\beta$~\cite{Verdu1998}. $\frac{{\mathbf{R}}}{M}$ is equivalent
to the covariance matrix of a system with equivalent system load
$\beta'=\frac{KL}{MN}=\frac{L}{M}\beta$. Thus as
$K,N\rightarrow\infty$, we have
\begin{eqnarray}
   \frac{\mbox{trace}\left\{\bf{\Sigma}_a\right\}}{M}\rightarrow
   \frac{\sigma_n^2}{M-L\beta}.\nonumber
\end{eqnarray}
Therefore, for sufficiently large $K$ and $N$, the variance of
channel estimation error is given by
\begin{eqnarray}
\Delta_a=\frac{\sigma_n^2}{M-L\beta},
\end{eqnarray}
which can be approximated by $\Delta_a\approx\frac{\sigma_n^2}{M}$
when $M$ is sufficiently large.

It should be noted that, in asynchronous systems, we can remove part
of the chips in the first and the last symbol periods to obtain a
similar matrix $\mathbf{S}_{NM-d_{max}\times KL}$, where $d_{max}$
denotes the largest time offsets of different users, measured in
chips. Since the training symbols have been incorporated into the
spreading codes, we can consider the columns of $\mathbf{S}$ as
random $(NM-d_{max})-$ vectors, regardless of the time offsets of
different users. Therefore, the variance of channel estimation error
in asynchronous systems is similar to that of synchronous systems
when $M$ is sufficiently large.

\subsection{Channel Estimation with Decision Feedback}
\subsubsection{Algorithm}
When decision feedback is used in place of training symbols to
derive the `ML' channel estimates\footnote{By `ML' estimates, we
mean using the expression obtained from the training symbol based
estimation, but with symbols obtained from decision feedback. It
is not an exact ML estimate since the distribution of the
decision feedback error is not considered.}, a process that
assumes that the decision feedback is free of error, the channel
estimation error is caused by both the thermal noise and the
decision feedback error. %Thus we can discuss the impact of each
%type of impairments separately.
On applying (\ref{MLestimation}),
the channel estimate with decision feedback is given by
\begin{eqnarray}
   \hat{\mathbf{a}}&=&\hat{\mathbf{R}}^{-1}\hat{\mathbf{y}}   \nonumber \\
          &=&\hat{\mathbf{R}}^{-1}\hat{\mathbf{S}}^T(\mathbf{S}\mathbf{a}+\mathbf{n})               \nonumber \\
          &=&\mathbf{a}+\hat{\mathbf{R}}^{-1}\hat{\mathbf{S}}^T(\delta\mathbf{S}\mathbf{a}+\mathbf{n}),       \nonumber
\end{eqnarray}
where $\delta\mathbf{S}\triangleq\mathbf{S}-\hat{\mathbf{S}}$,
$\hat{\mathbf{R}}\triangleq\hat{\mathbf{S}}^T\hat{\mathbf{S}}$,
$\hat{\mathbf{y}}\triangleq\hat{\mathbf{S}}^T{\mathbf{r}}$ and
$\hat{\mathbf{S}}$ is the version of ${\mathbf{S}}$ in
(\ref{MLestimation}) obtained from the decision feedback, which is
given by
\begin{eqnarray}
   &\hat{\mathbf{S}}&=\left(\left(\mathbf{S}(1)\hat{\mathbf{B}}(1)\right)^T,...,\left(\mathbf{S}(M)\hat{\mathbf{B}}(M)\right)^T\right)_{NM\times KL}^T,
   \nonumber\\
   &\hat{\mathbf{B}}(m)&=\left(
                     \begin{array}{cccc}
                       \hat{b}_1(m)\mathbf{I}_{L\times L} & 0 & \cdots & 0 \\
                       0 & \hat{b}_2(m)\mathbf{I}_{L\times L} & \cdots & 0 \\
                       \vdots & \vdots & \ddots & \vdots \\
                       0 & 0 & \cdots & \hat{b}_K(m)\mathbf{I}_{L\times L} \\
                     \end{array}
                   \right)_{KL\times KL}.\nonumber
\end{eqnarray}

Hence, the channel estimation error can be decomposed into two parts
\begin{eqnarray}
   \delta
   \mathbf{a}&=&-\hat{\mathbf{R}}^{-1}\hat{\mathbf{S}}^T(\delta\mathbf{S}\mathbf{a}+\mathbf{n})   \nonumber\\
           &=&\delta \mathbf{a}_f+\delta \mathbf{a}_n,
\end{eqnarray}
where
$\delta\mathbf{a}_f\triangleq-\hat{\mathbf{R}}^{-1}\hat{\mathbf{S}}^T\delta\mathbf{S}\mathbf{a}$
and $\delta \mathbf{a}_
n\triangleq-\hat{\mathbf{R}}^{-1}\hat{\mathbf{S}}^T\mathbf{n}$
denote the channel estimation error due to the decision feedback
error and the thermal noise, respectively. It is reasonable to
assume that $\delta\mathbf{a}_f$ and $\delta\mathbf{a}_n$ are
mutually independent. (Recall our assumption concerning the use of
only measurements $t\neq i$ in estimating gains at time $i$.)

It is difficult to tackle the calculation of $\delta\mathbf{a}$ due to the matrix
inversion ${\hat{\mathbf{R}}}^{-1}$. However, we can approximate
${\hat{\mathbf{R}}}^{-1}$ by $\frac{\mathbf{I}_{KL\times KL}}{M}$
when $P_e$ is sufficiently small. This approximation is justified by
the following lemma.
\begin{lem}
When fixing $K$ and $N$, we have
\begin{eqnarray}
M\hat{\mathbf{R}}^{-1}\rightarrow \mathbf{I}_{KL\times KL},\nonumber
\end{eqnarray}
almost surely \footnote{Here, a matrix is considered as a point in
the probability space and the metric is induced by a matrix norm.}
as $M\rightarrow\infty$ and $P_e\rightarrow 0$.
\end{lem}
\begin{proof}
According to the definition of $\hat{\mathbf{R}}$, we
have%\footnote{$\|\cdot\|_F$ denotes Frobenius norm of a matrix.}
\begin{eqnarray}
{\hat{\mathbf{R}}}^{-1}={\mathbf{R}}^{-1}+{\mathbf{R}}^{-1}\mathbf{A},\nonumber
\end{eqnarray}
where $\mathbf{A}=\left(\mathbf{I}-\delta
\mathbf{R}\mathbf{R}^{-1}\right)^{-1}-\mathbf{I}$. According to the
error analysis of matrix inversion in~\cite{Horm}, we
have\footnote{$x=O(P_e)$ means $\frac{x}{P_e}<\infty$ as
$P_e\rightarrow 0$.}
\begin{eqnarray}
E\left\{\|\mathbf{A}\|_F\right\}\leq E\left\{\frac{\|\delta
\mathbf{R}\mathbf{R}^{-1}\|_F}{1-\|\delta
\mathbf{R}\mathbf{R}^{-1}\|_F}\right\}=O(P_e),\nonumber
\end{eqnarray}
which tends to 0 as $P_e\rightarrow 0$. Thus, we have
\begin{eqnarray}
E\left\{\left\|{\hat{\mathbf{R}}}^{-1}-{\mathbf{R}}^{-1}\right\|_F\right\}
\leq
\left\|\mathbf{R}^{-1}\right\|_FE\left\{\left\|\mathbf{A}\right\|_F\right\}\rightarrow
0,\nonumber
\end{eqnarray}
as $P_e\rightarrow 0$. Therefore, ${\hat{\mathbf{R}}}^{-1}$
converges to ${\mathbf{R}}^{-1}$ almost surely as $P_e\rightarrow
0$.

Applying the strong law of large numbers and the fact that the
diagonal elements in $${\mathbf{R}}=\sum_{m=1}^M
(\hat{\mathbf{B}}(m)\mathbf{S}(m))^T\mathbf{S}(m)\hat{\mathbf{B}}(m)$$
are $M$ and the off-diagonal elements in
$(\hat{\mathbf{B}}(m)\mathbf{S}(m))^T\mathbf{S}(m)\hat{\mathbf{B}}(m)$
are independent for different values of $m$ and have zero mean, we
obtain that, while keeping $K$ and $N$ fixed,
$\frac{\mathbf{R}}{M}\rightarrow \mathbf{I}_{KL\times KL}$ almost
surely, as $M\rightarrow\infty$. Since the elements of
${\mathbf{R}}^{-1}$ are continuous functions of those in
${\mathbf{R}}$ in a neighborhood of
${\mathbf{R}}=M\mathbf{I}_{KL\times KL}$, we also have
$M{\mathbf{R}}^{-1}\rightarrow \mathbf{I}_{KL\times KL}$ as
$M\rightarrow\infty$. This completes the proof.
\end{proof}

Therefore, we can further approximate $\hat{\mathbf{R}}^{-1}$ by
$\frac{\mathbf{I}_{KL\times KL}}{M}$ for large $M$ and small
$P_e$. For simplicity, our further discussion of $\delta
\mathbf{a}_f$ will be based on this approximation, which will be
validated by numerical results. Consequently, in the following
discussions, we use the approximations
$$\delta
\mathbf{a}_f=-\frac{1}{M}\hat{\mathbf{S}}^T\delta\mathbf{S}\mathbf{a},$$
and
$$\delta
\mathbf{a}_n=-\frac{1}{M}\hat{\mathbf{S}}^T\mathbf{n}.$$

\subsubsection{Covariance matrix of channel estimation error}
We denote the covariance matrices of $\delta \mathbf{a}$, $\delta
\mathbf{a}_f$ and $\delta \mathbf{a}_n$ by $\bf{\Sigma}_a$,
$\bf{\Sigma}_f$ and $\bf{\Sigma}_n$, respectively, which satisfy
$\bf{\Sigma}_a=\bf{\Sigma}_f+\bf{\Sigma}_n$. We first consider the
channel estimation error incurred by decision feedback errors. The
following lemma shows that the channel estimation error $\delta
\mathbf{a}_f$ is asymptotically biased. The proof is given in
Appendix II.
\begin{lem}\label{Lem_Mean_a_f}
When keeping $K$ and $N$ fixed, we have
\begin{eqnarray}\label{Mean_a_f}
E\{\delta{\mathbf{a}}_f\}\rightarrow 2P_e\mathbf{a},
\end{eqnarray}
almost surely, as $M\rightarrow\infty$.
\end{lem}
It should be noted that this bias cannot be removed \textit{a
priori} in the estimator since it is dependent on the channel
gain, $\mathbf{a}$. However, this bias vanishes as
$P_e\rightarrow 0$.

An asymptotic expression for the elements in $\bf{\Sigma}_f$ is
given in the following proposition, whose proof is given in Appendix
III, where we also explain that the conclusion also applies to
asynchronous case when $P_e$ is sufficiently small.
\begin{prop}\label{Lemma_cov_f}
For all $i$ and $j$, when fixing $K$ and $N$, we have
that (recall that $a_{kl}$ is the channel gain of use $k$ and path
$l$)
\begin{eqnarray}\label{Cov_delta_a_f}
M\times\left(\bf{\Sigma}_f\right)_{ij}
&\rightarrow&\left\{\begin{array}{lll} 4P_e\left(|{a}_{\lceil
\frac{i}{L}\rceil,\mbox{ }mod(i,L)}|^2+\frac{1}{N}\sum_{k=1\mbox{,
}k\neq i}^{KL}|{a}_{\lceil
\frac{k}{L}\rceil,\mbox{ }mod(k,L)}|^2\right),\qquad \mbox{if  } i=j,\\
4P_e\left(1+\frac{1}{N}\right){a}_{\lceil \frac{i}{L}\rceil,\mbox{
}mod(i,L)}{a}_{\lceil \frac{j}{L}\rceil,\mbox{ }mod(j,L)}^*, \qquad
\mbox{if } i\neq j \mbox{ and }\lceil
\frac{i}{L}\rceil=\lceil \frac{j}{L}\rceil,\\
4P_e^2\left(1+\frac{1}{N}\right){a}_{\lceil \frac{i}{L}\rceil,\mbox{
}mod(i,L)}{a}_{\lceil \frac{j}{L}\rceil,\mbox{ }mod(j,L)}^*,
\qquad\mbox{if } \lceil \frac{i}{L}\rceil\neq\lceil
\frac{j}{L}\rceil
\end{array}\right.,
\end{eqnarray}
almost surely, as $M\rightarrow\infty$.
\end{prop}

For $\delta \mathbf{a}_n$, which is caused by thermal noise, the
corresponding analysis is identical to that of training symbol based
estimation. Then, we have
\begin{eqnarray}\label{Cov_delta_a_n}
   M\bf{\Sigma}_n&=&M\mbox{cov}\left(\hat{\mathbf{R}}^{-1}\hat{\mathbf{S}}^T\mathbf{n}\right)           \nonumber \\
           &=&M\sigma_n^2\hat{\mathbf{R}}^{-1}\nonumber\\
           &\rightarrow&\sigma_n^2 \mathbf{I}_{KL\times KL},
\end{eqnarray}
almost surely, as $M\rightarrow \infty$. Then the covariance matrix
of channel estimation error ${\bf{\Sigma}_a}\triangleq
E\left\{\delta \mathbf{a}\delta
\mathbf{a}^H\right\}=\bf{\Sigma}_f+\bf{\Sigma}_n$ can be obtained
from (\ref{Cov_delta_a_f}) and (\ref{Cov_delta_a_n}).

\subsubsection{Variance of channel estimation error}
The variance of channel estimation error can be obtained as a
corollary of the previous subsection.
\begin{cor}
On defining
$\Delta_{a}\triangleq\frac{1}{KL}\mbox{trace}\left\{\bf{\Sigma}_a\right\}$,
we have
\begin{eqnarray}
   M\Delta_{a}&\rightarrow&\frac{4P_e(1+\beta L)}{L}+\sigma_n^2,
\end{eqnarray}
almost surely, as $K,N,M\rightarrow\infty$.
\end{cor}

Thus, when $K,N,M$ are sufficiently large, we have the following
approximation
\begin{eqnarray}\label{Delta_a}
   \Delta_{a}&\approx&\frac{4P_e(1+\beta L)}{LM}+\frac{\sigma_n^2}{M}.
\end{eqnarray}

It should be noted that the channel estimation error cannot be
removed by increasing $M$ although the variance vanishes as
$M\rightarrow\infty$, since the estimate is biased and the bias
cannot be removed \textit{a priori}.

\subsection{Estimation with Both Training Symbols and Decision Feedback}
We denote the number of training symbols by $M_t$ and the
corresponding percentage by $\alpha=\frac{M_t}{M}$. When the
training symbols and decision feedback are combined for channel
estimation, the performance is determined by (\ref{Delta_a}), with
$P_e$ replaced by $(1-\alpha)P_e$.
Decision feedback should only be used along with the training symbols if
the resulting variance is smaller than that obtained when only
the training symbols are used.
Then it is easy to check that,
when $M$ and $M_t$ are sufficiently large, $P_{e\max}$, the
maximum $P_e$ assuring performance improvement when decision
feedback is used, is determined by
\begin{eqnarray}
\frac{4(1-\alpha)P_e(1+\beta L)}{LM}+\frac{\sigma_n^2}{M}
\leq \frac{\sigma_n^2}{M_t},\nonumber
\end{eqnarray}
which results in
\begin{eqnarray}
P_{e\max}=\frac{\sigma_n^2L}{4\alpha(1+\beta L)},
\end{eqnarray}
from which we observe that $P_{e\max}$ decreases with $\alpha$ and
$\beta$ while increasing with $\sigma_n^2$ and $L$.

\section{PIC and Channel Decoder}
\subsection{Performance Analysis of PIC}
For convenience of analysis, the performance of PIC is analyzed based on matched filter (MF) outputs.
We drop the index of the symbol period for notational simplicity throughout this section.
For a given symbol period, the MF outputs, which
form sufficient statistics for multiuser detection, are given
by
\begin{eqnarray}
    \mathbf{y}= \mathbf{S}^T\mathbf{r}\nonumber.
\end{eqnarray}
In PIC based
 multiuser detection, the MAI reconstructed from the
channel estimates and the decoder output is subtracted directly
from the MF output of the desired user. Without loss of
generality, we take the $l$-th path of user 1 as an example; then the MF
output after PIC, which is contaminated by residual MAI and
thermal noise $n_{1l}=\mathbf{s}_{1l}^T\mathbf{n}$, is given by
\begin{eqnarray}
y_{1l}={a}_{1l}b_1+\sum_{m\neq l}a_{1m}\rho_{1l1m}b_1+I_{1l},
\end{eqnarray}
where
$$I_{1l}=\sum_{k=2}^K\sum_{m=1}^L\rho_{1lkm}\left(a_{km}b_k-\hat{a}_{km}\hat{b}_k\right)+n_{1l},$$
which is the sum of the residual interference and the thermal
noise. It is obvious that $E\{I_{1l}\}=0$. And the corresponding
variance is given by
\begin{eqnarray}\label{PICOut}
\sigma_I^2\triangleq
E\left\{\left|I_{1l}\right|^2\right\}&=&\frac{1}{N}\sum_{k=2}^K\sum_{m=1}^LE\left\{\left|\delta
a_{km}b_k+\delta b_{k}a_{km}-\delta a_{km}\delta b_{k}\right|^2\right\}+E\left\{\left|n_{1l}\right|^2\right\}\nonumber\\
&=&\frac{1}{N}\sum_{k=2}^K\sum_{m=1}^L\bigg\{E\left\{\left|\delta
a_{km}\right|^2\right\}+4P_e\left|a_{km}\right|^2+4P_eE\left\{\left|\delta
a_{km}\right|^2\right\}
+2E\left\{\delta a_{km}a^*_{km}\right\}E\left\{b_k\delta b_k\right\}\nonumber\\
&&-2E\left\{\left|\delta a_{km}\right|^2\right\}E\left\{b_k\delta
b_k\right\}
-8P_eE\left\{a_{km}\delta a^*_{km}\right\}\bigg\}+\sigma_n^2\nonumber\\
&\rightarrow&\beta L\Delta_a+4\beta(1-P_e)P_e+\sigma_n^2,
\end{eqnarray}
as $K,N\rightarrow\infty$, where we have applied the fact that
$E\left\{\left|\delta
a_{km}\right|^2\right\}=\Delta_a+4P_e^2\left|a_{km}\right|^2$,
$E\left\{a_{km}\delta a^*_{km}\right\}=2P_e\left|a_{km}\right|^2$,
$E\left\{b_k\delta b_k\right\}=2P_e$. It is easy to check that
$\sigma_I^2$ is identical for asynchronous systems since different
time offsets do not affect the interference power.

It is difficult to apply the central limit theorem to show the
asymptotic normality of the PIC output since the variables
$\left\{\delta a_{km}\right\}$ are mutually correlated across
different users and paths. However, numerical results in Section
VI will show that the output distribution of PIC can be well
approximated by a Gaussian distribution. Thus, in the subsequent
sections, we assume that the output of PIC is Gaussian
distributed.

According to the properties of the crosscorrelation given in Section
II.A, $\rho_{1l1m}\rightarrow 0$ almost surely, as
$N\rightarrow\infty$. Thus, for large spreading gain, the
interference across different paths of the same user can be ignored.
With the normality assumption of the residual MAI, it is easy to
show that the variables $\left\{I_{1l}\right\}_{l=1,...,L}$ are
mutually independent as $N\rightarrow\infty$, which means that
channel coded symbol $b_1$ is transmitted through $L$ independent
channels. This assumption simplifies the analysis although it does
not hold exactly when $N$ is finite. Thus, we use MRC to collect
these $L$ replicas, resulting in the output
\begin{eqnarray}\label{PICOutput}
z_1=\sum_{l=1}^L\hat{a}^*_{1l}a_{1l}b_1+\sum_{l=1}^L\hat{a}_{1l}^*I_{1l}.
\end{eqnarray}

Applying Lemma \ref{Lem_Mean_a_f}, we obtain that, as
$M,L\rightarrow\infty$,
\begin{eqnarray}
\sum_{l=1}^L\hat{a}^*_{1l}a_{1l}b_1&=&\sum_{l=1}^L\left(\left|a_{1l}\right|^2-\delta
a_{1l}^*a_{1l}\right)\nonumber\\
&\rightarrow&1-\sum_{l=1}^\infty E\left\{\delta
a^*_{1l}\right\}a_{1l}\nonumber\\
&=&1-2P_e\sum_{l=1}^\infty\left|a_{1l}\right|^2\nonumber\\
&=&1-2P_e.\nonumber
\end{eqnarray}

Moreover, we can obtain that, as $M,L\rightarrow\infty$
\begin{eqnarray}
E\left\{\left|\sum_{l=1}^L\hat{a}_{1l}^*I_{1l}\right|^2\right\}&=&\sum_{l=1}^LE\left\{\left|\hat{a}_{1l}^*\right|^2\right\}\sigma_I^2\nonumber\\
&=&\left(1-2\sum_{l=1}^LE\left\{\delta
a^*_{1l}a_{1l}\right\}+\sum_{l=1}^LE\left\{\left|\delta
a^*_{1l}\right|^2\right\}\right)\sigma_I^2\nonumber\\
&\rightarrow&\left(1-4P_e+4P_e^2+L\Delta_a\right)\sigma_I^2\nonumber\\
&=&((1-2P_e)^2+L\Delta_a)\sigma_I^2.\nonumber
\end{eqnarray}

Therefore, when $M$ and $L$ are sufficiently large,
(\ref{PICOutput}) can be approximated by
\begin{eqnarray}\label{ApproxPICOut}
z_1\approx(1-2P_e)b_1+n_1,
\end{eqnarray}
where $n_1$ is a CSCG random variable with variance of
$((1-2P_e)^2+L\Delta_a)\sigma_I^2$. An interesting observation is
that the channel estimation error not only increases the
interference but also decreases the valid received power of the
desired user.

\subsection{Performance of Channel Decoder}
At the channel decoder, $P_e$ is a function of the input
signal-to-interference-plus-noise ratio (SINR) at the input to the
channel decoder given by
\begin{eqnarray}\label{Def_fun_g}
   P_e=g\left(\frac{1}{\mbox{SINR}}\right),
\end{eqnarray}
where the function $g$ can be estimated using Monte Carlo
simulations. For most practical channel codes, the following
assumption is reasonable:
\begin{assum}
Within a closed interval $\Omega=[0,\sigma_I^{max}]$, function $g$
satisfies
\begin{itemize}
\item $g(x)$ monotonically increases with $x$, and $g(0)=0$;

\item $g(x)$ is continuously differentiable and $g'(0)=0$.
\end{itemize}
\end{assum}

\section{Analysis of System Performance}
In this section, we analyze the overall iterative system shown in
Figure 1. We consider only the case of small $P_e$, moderate
$\sigma_n^2$ and moderate $M$ and note that the analytic results
become more precise as $P_e$ and $\sigma_n^2$ decrease and $M$
increases. This configuration is reasonable for the decision
feedback based%based
 systems since if $M$ is large, training symbol based
channel estimation can be adopted with marginal loss of spectral
efficiency; if $M$ is small, it is difficult to carry out coherent
detection; and if $P_e$ is large, the iteration diverges. Although
the performance analysis of the channel estimation in Section III is
based on large $M$, numerical results in Section VI indicate that
expression (\ref{Delta_a}) is still valid for moderate $M$. We adopt
the expressions (\ref{Delta_a}) and (\ref{PICOut}) in large system
limits ($K,N\rightarrow\infty$).

\subsection{Iterative Mapping}
In this section, we consider the $d$-th iteration and couple the
results from Section III and Section IV to analyze the overall
system performance. We can regard the decoding process as an
iterative mapping $h:\mathbb{R}\rightarrow \mathbb{R}$ in terms of
the error probability of the decoder output after the $d$-th iteration,
$P_e^{(d)}$, which is given by (recall that $g$ is defined as the
function characterizing the output error probability in terms in input SINR
in (\ref{Def_fun_g}))
\begin{eqnarray}\label{IterMap}
P_e^{(d)}&=&h(P_e^{(d-1)})\nonumber\\
&\approx&g\left(D_0+D_1P_e^{(d-1)}\right),
\end{eqnarray}
where we ignore terms of a smaller order than $P_e$ and
$\frac{1}{M}$ since we assume small $P_e$ and large (or moderate)
$M$. Based on (\ref{Delta_a}), (\ref{PICOut}) and
(\ref{ApproxPICOut}), the coefficients $D_0$ and $D_1$ are given by
\begin{eqnarray}
\left\{
\begin{array}{ll}
  &D_0=\sigma_n^2\left(1+\frac{\beta L}{M}+\frac{L\sigma_n^2}{M}\right) \\
  &D_1=4\left(\beta+\frac{\beta+\sigma_n^2\beta L^2+\beta^2L+\sigma_n^2 L+L\beta\sigma_n^2+L\left(\sigma_n^2\right)^2}{M}\right) \\
\end{array}%
 \right..\nonumber
\end{eqnarray}

\subsection{Condition for Convergence}
A reasonably good initialization, which results in sufficiently
small channel estimation error and MAI in the first iteration, is
necessary to guarantee the convergence of the iterative mapping
described in (\ref{IterMap}). In the initial stage, only
training symbols are used for the channel estimation since no
decision feedback is available then. Any non-iterative multiuser
detection technique can be applied to the initializing stage. For
practical applications, we can use the LMMSE detector, whose
performance using imperfect channel estimation can be obtained using
the replica method~\cite{LiHuEuroSip2005}.

For convergence, the variance of input interference and noise of the
initializing stage, denoted by $\sigma_I^2(0)$ and obtained from the
SINR of the LMMSE detector, must satisfy the following conditions:
\begin{itemize}
\item $\sigma_I^2(0)$ is located within the interval $\Omega$
defined in Section IV.B, namely
\begin{eqnarray}\label{Conv_con1}
   \sigma_I^2(0)< \sigma_I^{max}.
\end{eqnarray}
This condition assures a reasonably good initial performance of the
iterations.

\item The variance of interference and noise decreases with
iteration time, namely
\begin{eqnarray}\label{Conv_con2}
   g(\sigma_I^2(0))< \frac{\sigma_I^2(0)-D_0}{D_1}.
\end{eqnarray}
This condition assures that the iterations do not diverge.
\end{itemize}

\subsection{Condition Assuring the Uniqueness of the Fixed Point}
If there exists more than one fixed point, the iteration may
become stuck at a suboptimal fixed point and not converge to the
optimal one. The following proposition provides a sufficient
condition for the uniqueness of the fixed point and the
corresponding convergence rate.
\begin{prop}
(1) If there exists a $\gamma < 1$, such that
\begin{eqnarray}\label{Con_D1}
D_1\leq\frac{\gamma}{\max_{x\in \Omega}\left(g'(x)\right)},
\end{eqnarray}
then there exists only one fixed point $x_f$ for the iterative
mapping $x_{k+1}=h\left(x_k\right)$, and for every initial point
$x_0\in \Omega$, the mapping converges to $x_f$ with an
exponential rate, namely $\left\|x_k-x_f\right\|\leq
\frac{\gamma^k}{1-\gamma}\left\|x_0-x_f\right\|$.

(2) If there exists an $x_1\in\Omega$ such that
$\frac{1}{g'(x_1)}<D_1<\frac{x_1}{g(x_1)}$, then there exists a
$D_0$ such that there is more than one fixed point for $h$.
\end{prop}
\begin{proof}
(1) The condition $D_1\leq\frac{\gamma}{\max_{x\in
\Omega}\left(g'(x)\right)}$ implies that
$h'(x)=g'(D_0+D_1x)\leq\gamma<1$. Then $h(\cdot)$ is a contraction
mapping, and the conclusions follow due to Banach's fixed point
theorem~\cite{Kreyszig1998}.

(2) Letting $x_f=g(x_1)$ and setting $D_0=x_1-D_1x_f$, we can show
that $D_0>0$ due to the assumption that
$D_1<\frac{x_1}{g(x_1)}=\frac{x_1}{x_f}$. It is easy to check that
$x_f$ is a fixed point and $g'(D_0+D_1x_f)=D_1g'(x_1)>1$. Hence,
there exists an $\epsilon>0$ such that for all $x\in
\left(x_f,x_f+\epsilon\right)$, $g(D_0+D_1x)>x$. However,
$g(D_0+D_1x_2)<x_2$ for $x_2=g\left(\sigma_I^2(0)\right)$ due to
condition (\ref{Conv_con2}). If $x_2<x_f$, there exists at least one
fixed point within $(0,x_2)$ since $g(D_0)>0$; if $x_2>x_f$, there
exists at least one fixed point different from $x_f$ within
$(x_f,x_2)$.
\end{proof}

It should be noted that condition (\ref{Con_D1}) is sufficient but
not necessary for the uniqueness of the fixed point. This condition
is more stringent than the condition of convergence in
(\ref{Conv_con2}) since it assures both the uniqueness of the fixed
point and the exponential convergence rate. The second part shows
that a moderate $D_1$ may cause multiple fixed points. A useful
conclusion drawn from (\ref{Con_D1}) is that this iterative
procedure does not work well for those channel codes, such as
powerful turbo codes or LDPC codes, that have a steep
performance curve (bit error rate versus SINR) which implies a
large value of $\max_{x\in \Omega}\left(g'(x)\right)$.
This will be demonstrated in numerical
simulations in Section VI.

\subsection{Asymptotic Multiuser Efficiency}
As is described in~\cite{Verdu1998}, the asymptotic multiuser efficiency
measures the slope at which the bit-error-rate goes to zero in logarithmic scale,
giving intuition into the performance loss from multiuser interference.

Suppose that there is only one fixed point for the iterative mapping
$h$, and let $P_e(\sigma_n^2)$ be this fixed point when the noise
power is $\sigma_n^2$.  Similarly, let $D_0(\sigma_n^2)$ and
$D_1(\sigma_n^2)$ be the corresponding values of $D_0$ and $D_1$ in
(\ref{IterMap}). It is obvious that $P_e(0)=0$ and $D_0(0)=0$.

The asymptotic multiuser efficiency 
 is given by
\begin{eqnarray}
   \mbox{AME}&=&\lim_{\sigma_n^2\rightarrow 0}\frac{\sigma_n^2}{D_0(\sigma_n^2)+D_1(\sigma_n^2)P_e(\sigma_n^2)}   \nonumber \\
      &=&\frac{1}{\frac{dD_0(\sigma_n^2)}{d\sigma_n^2}\bigg|_{\sigma_n^2=0}+\frac{d(D_1(\sigma_n^2)P_e(\sigma_n^2))}{d\sigma_n^2}\bigg|_{\sigma_n^2=0}}
      \nonumber.
\end{eqnarray}
If
$H(P_e,\sigma_n^2)=g\left(D_0(\sigma_n^2)+D_1(\sigma_n^2)P_e\right)-P_e$,
then $P_e(\sigma_n^2)$ is the unique solution of
$H(P_e,\sigma_n^2)=0$. Applying the assumptions that $g'(0)=0$ and
$P_e(0)=0$, we have
\begin{eqnarray}
    \frac{d(D_1(\sigma_n^2)P_e(\sigma_n^2))}{d\sigma_n^2}\bigg|_{\sigma_n^2=0}
    &=&D_1(0)\frac{dP_e(\sigma_n^2)}{d\sigma_n^2}\bigg|_{\sigma_n^2=0}    \nonumber \\
    &=&-D_1(0)\frac{\frac{\partial H(P_e,\sigma_n^2)}{\partial \sigma_n^2}\bigg|_{\sigma_n^2=0}}{\frac{\partial H(P_e,\sigma_n^2)}{\partial    P_e}\bigg|_{\sigma_n^2=0}}\nonumber \\
    &=&-D_1(0)\frac{\frac{\partial (D_0(\sigma_n^2)+D_1(\sigma_n^2)P_e)}{\partial\sigma_n^2}\bigg|_{\sigma_n^2=0}g'(0)}{D_1(0)g'(0)-1}\nonumber \\
    &=&0 \nonumber.
\end{eqnarray}
Thus
\begin{eqnarray}\label{AME}
   \mbox{AME}&=&\frac{1}{\frac{dD_0(\sigma_n^2)}{d\sigma_n^2}|_{\sigma_n^2=0}}\nonumber\\
             &=&\frac{1}{1+\frac{L\beta}{M}}.
\end{eqnarray}
From (\ref{AME}), we can see that the loss of AME is due to the channel
estimation error incurred by the thermal noise. The impact of the
decision feedback error vanishes as $\sigma_n^2\rightarrow 0$,
while that of the channel estimation error remains.
\subsection{Computational Aspect}
The main computational cost of the iterative channel estimation and
multiuser detection includes:
\begin{itemize}
\item Solving the linear equation $\hat{\mathbf{R}}\hat{\mathbf{a}}=\mathbf{y}$ for ML channel estimation.
\item Reconstructing the channel symbols and cancelling the interference.
\item Channel decoding.
\end{itemize}
Since the channel symbol reconstruction is similar to the encoding
procedure and the interference cancelation requires only
subtractions, this is not a bottleneck of the whole procedure and
the corresponding computational cost is of complexity $O(K)$.
Real-time channel decoding can also be accomplished in a way similar
to Turbo codes. Therefore, the main bottleneck is solving the
linear equation for channel estimation.

Direct Gaussian Eliminatation, which is of complexity $O(K^3)$, can
be applied to solve the equation
$\hat{\mathbf{R}}\hat{\mathbf{a}}=\mathbf{y}$ when $K$ is small.
When $K$ is large, iterative techniques of solving linear equations,
such as the Jacobi method and the Gauss-Seidel method, can be
applied. For assuring the convergence, we cite the following lemma
from \cite{Golub}:
\begin{lem}
The sufficient and necessary condition for the convergence of iterations
in solving the linear equation $\mathbf{A}\mathbf{x}=\mathbf{y}$ is that
\begin{itemize}
\item $\mathbf{A}$ and $2\mbox{ diag}(\mathbf{A})-\mathbf{A}$ are
both positive definite in the Jacobi
method\footnote{diag$(\mathbf{X})$ denotes a diagonal matrix
constituted by the diagonal elements in matrix $\mathbf{X}$};
\item $\mathbf{A}$ is positive definite in the Gauss-Seidel method.
\end{itemize}
\end{lem}

The Gauss-Seidel method always converges when $\beta<1$ since
$\hat{\mathbf{R}}$ is positive definite when $K<N$. For the Jacobi
method, it is easy to check that
$\mbox{diag}(\hat{\mathbf{R}})=\mathbf{I}_{K\times K}$. Since the
largest eigenvalue of $\hat{\mathbf{R}}$ converges to
$\left(1+\sqrt{\beta}\right)^2$ \cite{Bai1993} almost surely as
$K,N\rightarrow\infty$, the eigenvalues in
$2\mbox{ diag}(\hat{\mathbf{R}})-\hat{\mathbf{R}}$ are less than
$2-\left(1+\sqrt{\beta}\right)^2$ almost surely in the large system
limit. Therefore, $\sqrt{\beta}<1$ is a sufficient condition for the
almost sure convergence of Jacobi iteration in the large system
limit. Then, when $K$ and $N$ are sufficiently large and $K<N$, we
can use either Gauss-Seidel or Jacobi iterations to estimate the
channel coefficients efficiently.

\section{Numerical Results}
\subsection{Channel Estimation}
Figure 2 shows the average variance of the channel estimates versus
the coherence time $M$ with the configuration of $\beta=0.2$, $L=5$,
$M_t=0$, $P_e=0.1$ and the signal-to-noise ratio
(SNR)$=5$dB\footnote{Note that $P_e$ and SNR are not mutually
independent; however, we set these two parameters arbitrarily to test
the validity of asymptotic results.}. The asymptotic results
obtained from (\ref{Delta_a}) and the simulation results for finite
systems ($N=100$) with spreading codes for the shifted model are
represented by solid and dotted curves, respectively. In this
figure, the estimation error variance caused by decision feedback
and noise are denoted by $\Delta_f$ and $\Delta_n$, respectively.
The corresponding asymptotic results are obtained from the first and
the second terms in (\ref{Delta_a}), respectively. We can observe
that the asymptotic results match the simulation results well even
when $M$ is small. This figure also demonstrates the validity of
results based on the independence assumption of the spreading codes
given in Section II.A.

\subsection{Normality of PIC Output}
Figure 3 shows the channel symbol error rate\footnote{This channel
symbol error rate is equivalent to bit error rate when the output of
PIC is used directly for the detection (without channel decoding).}
with the configuration of $\mbox{SNR}=10$dB, $K=N=30$ and $P_e=0.1$
and $0.05$. The solid curves represent the results obtained from
numerical simulations and the dashed curves represent the results
with the assumption that the output of PIC is CSCG distributed. The
gap between the numerical results and CSCG based prediction is
small, thus justifying the normality assumption of the PIC output.

\subsection{User Capacity}
We define the user capacity to be the maximum system load
$\beta_{max}$ with which the system can achieve the information bit
error rate of $10^{-3}$. Two types of channel codes, the
convolutional code $(35,23)_8$ and a turbo code (with two
constituent codes $(37,21)_8$), with bit rate $R=\frac{1}{2}$ and
codeword length 1024 are used in this paper and their error rates
for both information bits and extrinsic information based channel
symbols are shown in Figure 4. The corresponding $\beta_{max}$'s for
various values of coherence time $M$, denoted by `iterative', are
given in Figure 5 and Figure 6 for convolutional codes and turbo
codes, respectively, with the configuration $\alpha=0.2$, SNR$=5$dB
and $L=5$. The $\beta_{max}$'s of the non-iterative LMMSE detector,
denoted by `LMMSE', are given for comparison. We can see that the
iterative system achieves substantially higher user capacity than
the non-iterative one. The performance of systems with ideal
initialization, where actual channel parameters are provided by a
genie in the initialization stage, denoted by `Perfect
initialization', implies that a good initialization can improve the
performance considerably. Thus, blind or semi-blind non-iterative
techniques, which make use of information symbols, can be applied to
obtain a better initialization. For comparison, the user capacities
of both iterative and non-iterative systems with perfect channel
state information are also given in both figures. An interesting
observation is that the relative performance gain of iterative
systems over the non-iterative ones is smaller for turbo codes than
for convolutional codes. This is due to the steeper waterfall region
in turbo codes.

\section{Conclusions}
In this paper, we have analyzed the performance of decision feedback
based iterative channel estimation and multiuser detection in
multipath DS-CDMA channels. The decoding process has been described as
an iterative mapping in terms of the variance of the channel decoder
output, and conditions assuring the convergence and uniqueness
of a fixed point have been proposed. Numerical results show that the
initialization is important to the iterations, thus necessitating
the use of non-iterative blind or semi-blind channel estimation
algorithms for initialization purposes. Another observation of interest is that the gain of
the iterative process over a non-iterative one is small when a
near-optimal channel coding scheme is used.

\appendices
\section{Validity of Independence Model for Spreading Codes}
In (1), for different values of $l$ and $m$, $\mathbf{s}_{kl}$ and
$\mathbf{s}_{km}$ are generated by the same binary sequence with
different offsets. Our purpose is to show that if $K$ and $N$ are
large enough, we can regard the shifted spreading codes of
different paths of a given user as independent sequences. The
properties based on this assumption, which are used for the system
performance analysis in this paper, include:
\begin{itemize}
\item The properties of crosscorrelation $\rho_{klmn}$ in Section
II.A.

\item The distribution of the eigenvalues of the matrix
$\mathbf{S}\mathbf{S}^T$, when developing the expression of
$\Delta_n$ for finite $M$ and large $K$ in Section III.C. Our
assumption means that the corresponding distribution of the
shifted model is asymptotically identical to that of the
independent model.
\end{itemize}

It is easy to check the first item using the symmetry of the binary
distribution. However, the validity of the second one is
non-trivial and is of considerable importance when applying the
theory of large random matrices to multipath fading channels. We
can tackle this problem by showing that the moments of the
eigenvalues in both models are the same via the following lemma.

\begin{lem}
Denote a generic eigenvalue of $\mathbf{S}\mathbf{S}^T$ by $\lambda$.
Then the $m$-th moment of $\lambda$ in the shifted model is given
by
$$
E\left\{\lambda^m\right\}=\sum_{k=1}^m\left(\beta'\right)^k\sum_{m_1+...+m_k=m}c(m_1,...,m_k),\qquad\mbox{as
}K\rightarrow\infty ,
$$
which is the same expression of that of the independent model, and
where the definition of $c(m_1,...,m_k)$ is given in
~\cite{Linbo2001} and $\beta'=\frac{LK}{MN}$.
\end{lem}
\begin{proof}
Using similar arguments to those in~\cite{Linbo2001}, we have
\begin{eqnarray}\label{ExpectEigValue}
&&\frac{1}{N}E\left\{\mbox{trace}\{(\mathbf{S}\mathbf{S}^T)^m\}\right\}\nonumber\\
&=&\frac{1}{N^{m+1}}\sum_{i_1,...,i_m=1}^K\sum_{j_1,...,j_m=1}^NE\{V_{i_m,j_1}V_{i_1,j_1}...V_{i_{m-1},j_m}V_{i_m,j_m}\},
\end{eqnarray}
where $V_{i,j}=\sqrt{N}\mathbf{S}_{ij}$.

For any $i_r\neq i_s$, $V_{i_r,j_p}=V_{i_s,j_q}$ when $\lceil
\frac{i_r}{L}\rceil=\lceil \frac{i_s}{L}\rceil$ and $j_p-j_q$
equals the offset difference between these two shifted sequences.
However, the probability of such events vanishes as
$K\rightarrow\infty$ since
\begin{eqnarray}
P\left(|i_r-i_s|<L\right)\leq\left(
\begin{array}{c}
  m \\
  2 \\
\end{array}\right)\frac{2L+1}{KL}\rightarrow 0,\qquad \mbox{as  } K\rightarrow
\infty.\nonumber
\end{eqnarray}
Thus, as $K\rightarrow\infty$, the term involving $V_{i,j}$'s of
different users, which are mutually independent, dominates the
summation in (\ref{ExpectEigValue}). The remaining part of the proof is the same as in
~\cite{Linbo2001}.
\end{proof}

The following lemma (Theorem 30.1 in~\cite{Billingsley}) provides
a sufficient condition for the equality of two probability
measures when their moments are identical .
\begin{lem}
Let $\mu$ be a probability measure on the real line having finite
moments $\alpha_k=\int_{-\infty}^{\infty}x^k\mu(dx)$ of all orders.
If the power series $\sum_{k=1}^\infty \alpha_k\frac{r^k}{k!}$ has
a positive radius of convergence, then $\mu$ is the only
probability measure with the moments
$\left\{\alpha_m\right\}_{m=1,2,...}$.
\end{lem}

For applying Lemma I.2, we need the following lemma which provides an upper bound for the
moments of the eigenvalues.
\begin{lem}
For any eigenvalue $\lambda$ of $\mathbf{S}\mathbf{S}^T$, there
exists a constant $C>\max(1,\beta')$ such that for $m=1,2,...$
\begin{eqnarray}\label{IeqEigenExp}
E\left\{\lambda^m\right\}<C^mm^{m-2}.
\end{eqnarray}
\end{lem}
\begin{proof}
The result follows by induction on $m$.

It is easy to verify that (\ref{IeqEigenExp}) holds when $m=1,2$.
Suppose $E\left\{\lambda^n\right\}<C^nn^{n-2}$, for $n=1,2,...,m$.
Use the following recursive formula~\cite{Linbo2001} to evaluate
$E\left\{\lambda^{m+1}\right\}$, which is given by
\begin{eqnarray}
E\left\{\lambda^{m+1}\right\}=\sum_{k=1}^{m+1}\beta'\sum_{m_1+...+m_k=m+1}E\left\{\lambda^{m_1-1}\right\}\cdots E\left\{\lambda^{m_k-1}\right\}.\nonumber
\end{eqnarray}

Then we have
\begin{eqnarray}
E\left\{\lambda^{m+1}\right\}&=&\beta'\left(1+mE\{\lambda\}+E\{\lambda^m\}+\sum_{k=2}^{m-1}\sum_{m_1+...+m_k=m+1}E\left\{\lambda^{m_1-1}\right\}...E\left\{\lambda^{m_k-1}\right\}\right)\nonumber\\
                             &<&\beta'\left(1+m\beta'+C^mm^{m-2}+\sum_{k=2}^{m-1}\sum_{m_1+...+m_k=m+1}\prod_{i=1}^kC^{m_i-1}m_i^{m_i-3}\right)\nonumber\\
                             &<&\beta'\left(1+m\beta'+C^mm^{m-2}+\sum_{k=2}^{m-1}\sum_{m_1+...+m_k=m+1}C^{m+1-k}m^{m-1-k}\right)\nonumber\\
                             &<&C^{m+1}\left(1+m^{m-1}+\sum_{k=2}^{m-1}\left(\begin{array}{ll}m\\k-1\end{array}\right)m^{m-1-k}\right)\nonumber\\
                             &<&C^{m+1}\left(1+m^{m-1}+\sum_{k=1}^{m-2}\left(\begin{array}{ll}m-1\\k\end{array}\right)m^{m-1-k}\right)\nonumber\\
                             &=&C^{m+1}\sum_{k=0}^{m-1}\left(\begin{array}{ll}m-1\\k\end{array}\right)m^{m-1-k}\nonumber\\
                             &=&C^{m+1}(1+m)^{m-1},\nonumber
\end{eqnarray}
where the first inequality is based the assumption on $n=1,...,m$
and the fact that $E\{\lambda\}=\beta'$; the third inequality
applies the condition that $C>\max(1,\beta')$ and
$m^{m-1}>m^{m-2}+m$ for $m>2$. This concludes the proof.
\end{proof}

Applying Stirling's formula and Lemmas I.1,2,3, we can obtain the
conclusion that the eigenvalue distribution of
$\mathbf{S}\mathbf{S}^T$ in the shifted model is identical to that
of the independent model, thus assuring the assumption that the
columns of $\mathbf{S}$ can be regarded as independent in the
large system limit.

\section{Proof of Lemma \ref{Lem_Mean_a_f}}
\begin{proof}
From the definition of $\delta{\mathbf{a}}_f$, we have
\begin{eqnarray}
E\{\delta{\mathbf{a}}_f\}=-\frac{1}{M}\left(E\{\mathbf{S}^T\delta\mathbf{S}\mathbf{a}\}-E\{\delta\mathbf{S}^T\delta\mathbf{S}\mathbf{a}\}\right).
\end{eqnarray}

We consider the term
$E\{\delta\mathbf{S}^T\delta\mathbf{S}\mathbf{a}\}$ first. It is
easy to check that (recall that $\mathbf{s}_{kl}$ denotes the
spreading code of user $k$ along path $l$)
\begin{eqnarray}
\frac{1}{M}E\left\{\left(\delta\mathbf{S}^T\delta\mathbf{S}\right)_{ij}\right\}&=&\frac{1}{M}\sum_{m=1}^M\mathbf{s}^T_{pq}(m)\mathbf{s}_{rs}(m)E\left\{\delta
b_p\delta b_r\right\}\nonumber \\
&=&\left\{
\begin{array}{ll}
0,\qquad\mbox{if }p\neq r\\
\frac{4P_e}{M}\sum_{m=1}^M\mathbf{s}^T_{pq}(m)\mathbf{s}_{rs}(m),\qquad
\mbox{if }p=r
\end{array}
\right.,\nonumber
\end{eqnarray}
where $p=\left\lceil\frac{i}{L}\right\rceil$,
$q=\mbox{mod}\left(i,L\right)$,
$r=\left\lceil\frac{j}{L}\right\rceil$,
$s=\mbox{mod}\left(j,L\right)$. It should be noted we applied the
fact that $E\left\{\delta b_p\delta b_r\right\}=4P_e$ in the second
equality.

According to Assumption \ref{Assum3}, the spread codes are mutually
independent for different users or different paths. Thus, by
applying the strong law of large numbers, we have
\begin{eqnarray}
\frac{1}{M}\sum_{m=1}^M\mathbf{s}^T_{pq}(m)\mathbf{s}_{rs}(m)\rightarrow
\left\{
\begin{array}{ll}
0,\qquad\mbox{if }(p,q)\neq(r,s)\\
1,\qquad \mbox{if }(p,q)=(r,s)
\end{array}
\right..\nonumber
\end{eqnarray}

Therefore, we have
\begin{eqnarray}
\frac{1}{M}E\left\{\left(\delta\mathbf{S}^T\delta\mathbf{S}\right)_{ij}\right\}
\rightarrow\left\{
\begin{array}{ll}
0,\qquad\mbox{if }i\neq j\\
\frac{4P_e}{M},\qquad \mbox{if }i=j
\end{array}
\right.,\qquad\mbox{almost surely, as }M\rightarrow\infty\nonumber
\end{eqnarray}

Similarly, we can show that
\begin{eqnarray}
\frac{1}{M}E\left\{\left(\mathbf{S}^T\delta\mathbf{S}\right)_{ij}\right\}
\rightarrow\left\{
\begin{array}{ll}
0,\qquad\mbox{if }i\neq j\\
\frac{2P_e}{M},\qquad \mbox{if }i=j
\end{array}
\right.,\qquad\mbox{almost surely, as }M\rightarrow\infty\nonumber
\end{eqnarray}

This completes the proof.
\end{proof}

\section{Proof of Prop. \ref{Lemma_cov_f}}
\begin{proof}
The covariance matrix $\mathbf{\Sigma}_f$ is given by
\begin{eqnarray}\label{Cov_f}
   \bf{\Sigma}_f&\triangleq&\frac{1}{M^2}\mbox{cov}\left(\hat{\mathbf{S}}^T\delta\mathbf{S}\mathbf{a}\right)          \nonumber \\
           &=&\frac{1}{M^2}E\left\{{\mathbf{S}}^T\delta\mathbf{S}\mathbf{a}\mathbf{a}^H
           \delta\mathbf{S}^T{\mathbf{S}}\right\}-\frac{1}{M^2}E\left\{{\mathbf{S}}^T\delta\mathbf{S}\mathbf{a}\mathbf{a}^H
           \delta\mathbf{S}^T{\delta\mathbf{S}}\right\}\nonumber\\
           &-&\frac{1}{M^2}E\left\{\delta{\mathbf{S}}^T\delta\mathbf{S}\mathbf{a}\mathbf{a}^H
           {\delta\mathbf{S}}^T\mathbf{S}^T\right\}
           +\frac{1}{M^2}E\left\{\delta{\mathbf{S}}^T\delta\mathbf{S}\mathbf{a}\mathbf{a}^H
           \delta\mathbf{S}^T\delta{\mathbf{S}}\right\}\nonumber\\
           &-&E\{\delta{\mathbf{a}}_f\}E\{\delta{\mathbf{a}}_f\}^H.
\end{eqnarray}

The elements in ${\mathbf{S}}^T\delta\mathbf{S}
\mathbf{a}\mathbf{a}^H \delta\mathbf{S}^T{\mathbf{S}}$ are given by
\begin{eqnarray}
   \left({\mathbf{S}}^T\delta\mathbf{S}\mathbf{a}\mathbf{a}^H\delta\mathbf{S}^T{\mathbf{S}}\right)_{ij}
   =\sum_{p=1}^{M}\sum_{q=1}^{M}\sum_{k=1}^{KL}\sum_{l=1}^{KL}{\tilde{\mathbf{s}}}_i^T(p)\delta \tilde{\mathbf{s}}_k(p){\tilde{\mathbf{s}}}_j^T(q)\delta
   {\tilde{\mathbf{s}}}_l(q)\textmd{a}_k\textmd{a}_l^{*},\nonumber
\end{eqnarray}
where ${\tilde{\mathbf{s}}}_i(p)\triangleq b_{\lceil
\frac{i}{L}\rceil}(p)\mathbf{s}_{\begin{scriptsize}\lceil
\frac{i}{L}\rceil,\mbox{ mod}(i,L)\end{scriptsize}}(p)$, namely the
spreading code (incorporating the channel symbol) of the $\mbox{
mod}(i,L)$-th path of user $\lceil \frac{i}{L}\rceil$ at symbol
period $p$, $\delta{\tilde{\mathbf{s}}}_i(p)\triangleq \delta
b_{\lceil \frac{i}{L}\rceil}(p)\mathbf{s}_{\begin{scriptsize}\lceil
\frac{i}{L}\rceil,\mbox{ mod}(i,L)\end{scriptsize}}(p)$ and
$\textmd{a}_k$ is the $k$-th element of vector $\mathbf{a}$ and
equals $a_{\left\lceil\frac{k}{L}\right\rceil,mod(k,L)}$. To compute
the corresponding expectation, we apply the following properties,
which are based on Assumption \ref{Assum4}:
\begin{itemize}
\item When $p=q$, if $\lceil \frac{k}{L}\rceil=\lceil
\frac{l}{L}\rceil$, $P(\delta \tilde{\mathbf{s}}_k(p)\neq 0, \delta
\tilde{\mathbf{s}}_l(q)\neq 0)=P_e$, since $\delta
\tilde{\mathbf{s}}_k(p)$ and $\delta \tilde{\mathbf{s}}_l(p)$ are
determined by the same decision feedback;

\item When $p=q$, if $\lceil \frac{k}{L}\rceil\neq\lceil
\frac{l}{L}\rceil$, $P(\delta \tilde{\mathbf{s}}_k(p)\neq 0, \delta
\tilde{\mathbf{s}}_l(q)\neq 0)=P_e^2$, since $\delta
\tilde{\mathbf{s}}_k(p)$ and $\delta \tilde{\mathbf{s}}_l(p)$ are
determined by decision feedback from different users;

\item When $p\neq q$, $P(\delta \tilde{\mathbf{s}}_k(p)\neq 0, \delta
\tilde{\mathbf{s}}_l(q)\neq 0)=P_e^2$, since $\delta
\tilde{\mathbf{s}}_k(p)$ and $\delta \tilde{\mathbf{s}}_l(p)$ are
determined by decision feedback from different symbol periods;

\item When $\delta \tilde{\mathbf{s}}_k(p)\neq 0$, $\delta
\tilde{\mathbf{s}}_k(p)=2{\tilde{\mathbf{s}}}_k(p)$.
\end{itemize}

Thus the expectation of $i-j$th element of
${\mathbf{S}}^T\delta\mathbf{S}\mathbf{a}\mathbf{a}^H\delta\mathbf{S}^T{\mathbf{S}}$
is given by
\begin{eqnarray}\label{sum_T1_T2_T3}
   &&E\left\{\left({\mathbf{S}}^T\delta\mathbf{S}\mathbf{a}\mathbf{a}^H\delta\mathbf{S}^T{\mathbf{S}}\right)_{ij}\right\}\nonumber\\
   &=&4P_e\sum_{p=1}^{M}\sum_{k=1}^{KL}\sum_{\lceil \frac{l}{L}\rceil=\lceil
\frac{k}{L}\rceil}{\tilde{\mathbf{s}}}_i^T(p)
{\tilde{\mathbf{s}}}_k(p){\tilde{\mathbf{s}}}_j^T(p)
   {\tilde{\mathbf{s}}}_l(p)\textmd{a}_k\textmd{a}_l^{*}\nonumber\\
   &+&4P_e^2\sum_{p=1}^{M}\sum_{k=1}^{KL}\sum_{\lceil \frac{l}{L}\rceil\neq\lceil
\frac{k}{L}\rceil}{\tilde{\mathbf{s}}}_i^T(p)
{\tilde{\mathbf{s}}}_k(p){\tilde{\mathbf{s}}}_j^T(p)
   {\tilde{\mathbf{s}}}_l(p)\textmd{a}_k\textmd{a}_l^{*}\nonumber\\
   &+&4P_e^2\sum_{\begin{scriptsize}\begin{array}{ll}p,q=1 \\ p\neq
   q\end{array}\end{scriptsize}}^{M}\sum_{k=1}^{KL}\sum_{l=1}^{KL}{\tilde{\mathbf{s}}}_i^T(p) {\tilde{\mathbf{s}}}_k(p){\tilde{\mathbf{s}}}_j^T(q)
   {\tilde{\mathbf{s}}}_l(q)\textmd{a}_k\textmd{a}_l^{*}\nonumber\\
   &=&T_1+T_2+T_3\nonumber,
\end{eqnarray}
where $T_1$, $T_2$ and $T_3$ represent the corresponding three
summations, respectively.

Applying the strong law of large numbers and the assumption on the
spreading codes that $\left\{{\tilde{\mathbf{s}}}_i(p)\right\}$ are
independent for different values of $i$ or $p$, we can obtain that,
as $M\rightarrow\infty$, the following conclusions hold almost
surely:
\begin{eqnarray}
\frac{1}{M}T_1&\rightarrow&\left\{\begin{array}{lll}
4P_e\left(|\textmd{a}_i|^2+\frac{1}{N}\sum_{k=1\mbox{, }k\neq
i}^{KL}|\textmd{a}_k|^2\right),\qquad \mbox{if  } i=j,\\
4P_e\left(1+\frac{1}{N}\right)\textmd{a}_i\textmd{a}_j^*, \qquad
\mbox{if } i\neq j \mbox{ and }\lceil
\frac{i}{L}\rceil=\lceil \frac{j}{L}\rceil,\\
0, \qquad\mbox{if } \lceil \frac{i}{L}\rceil\neq\lceil
\frac{j}{L}\rceil
\end{array}\right.\nonumber\\
\frac{1}{M}T_2&\rightarrow&\left\{\begin{array}{ll}
4P_e^2\left(1+\frac{1}{N}\right)\textmd{a}_i\textmd{a}_j^*,
\qquad\mbox{if } \lceil
\frac{i}{L}\rceil\neq\lceil \frac{j}{L}\rceil,\\
0, \qquad\mbox{if } \lceil \frac{i}{L}\rceil=\lceil
\frac{j}{L}\rceil
\end{array}\right.\nonumber\\
\frac{1}{M^2}T_3&\rightarrow&4P_e^2\textmd{a}_i\textmd{a}_j^*,\qquad
\forall i,j.\nonumber
\end{eqnarray}

We can apply the same manipulation and obtain that
$E\left\{{\mathbf{S}}^T\delta\mathbf{S}\mathbf{a}\mathbf{a}^H
\delta\mathbf{S}^T{\delta\mathbf{S}}\right\}=E\left\{\delta{\mathbf{S}}^T\delta\mathbf{S}\mathbf{a}\mathbf{a}^H
{\delta\mathbf{S}}^T\mathbf{S}\right\}=\frac{1}{2}E\left\{\delta{\mathbf{S}}^T\delta\mathbf{S}\mathbf{a}\mathbf{a}^H\delta\mathbf{S}^T\delta{\mathbf{S}}\right\}$
as $M\rightarrow\infty$. Therefore, we can obtain
(\ref{Cov_delta_a_f}) since the sum of the middle three terms in
(\ref{Cov_f}) is zero and $T_3$ cancels
$E\{\delta{\mathbf{a}}_f\}E\{\delta{\mathbf{a}}_f\}^H$.
\end{proof}

It should be noted that the above analysis is also valid for
asynchronous case when $P_e$ is sufficiently small. Similar to the
discussion in Section III.A, we can remove part of the chips in the
first and the last symbol periods to obtain a similar matrix
$\mathbf{S}_{NM-d_{max}\times KL}$, where $d_{max}$ denotes the
largest time offsets of different users, measured in chips. When
$P_e$ is sufficiently small and $M$ is sufficiently large, we can
ignore the terms scaled by $P_e^2$ and the edge effect in the first
and last symbol period. Then, we have
\begin{eqnarray}\label{sum_T1_T2_T3}
   E\left\{\left({\mathbf{S}}^T\delta\mathbf{S}\mathbf{a}\mathbf{a}^H\delta\mathbf{S}^T{\mathbf{S}}\right)_{ij}\right\}
   \approx 4P_e\sum_{k=1}^{KL}\sum_{\lceil \frac{l}{L}\rceil=\lceil
\frac{k}{L}\rceil}{\tilde{\mathbf{s}}}_i^T
{\tilde{\mathbf{s}}}_k{\tilde{\mathbf{s}}}_j^T
   {\tilde{\mathbf{s}}}_l\textmd{a}_k\textmd{a}_l^{*},\nonumber
\end{eqnarray}
where $\tilde{\mathbf{s}}_k$ is the $k$-th column of matrix
$\mathbf{S}$, which converges to $T_1$ as $M\rightarrow\infty$.

\newpage
\begin{figure}
  \centering
  \includegraphics[scale=0.6]{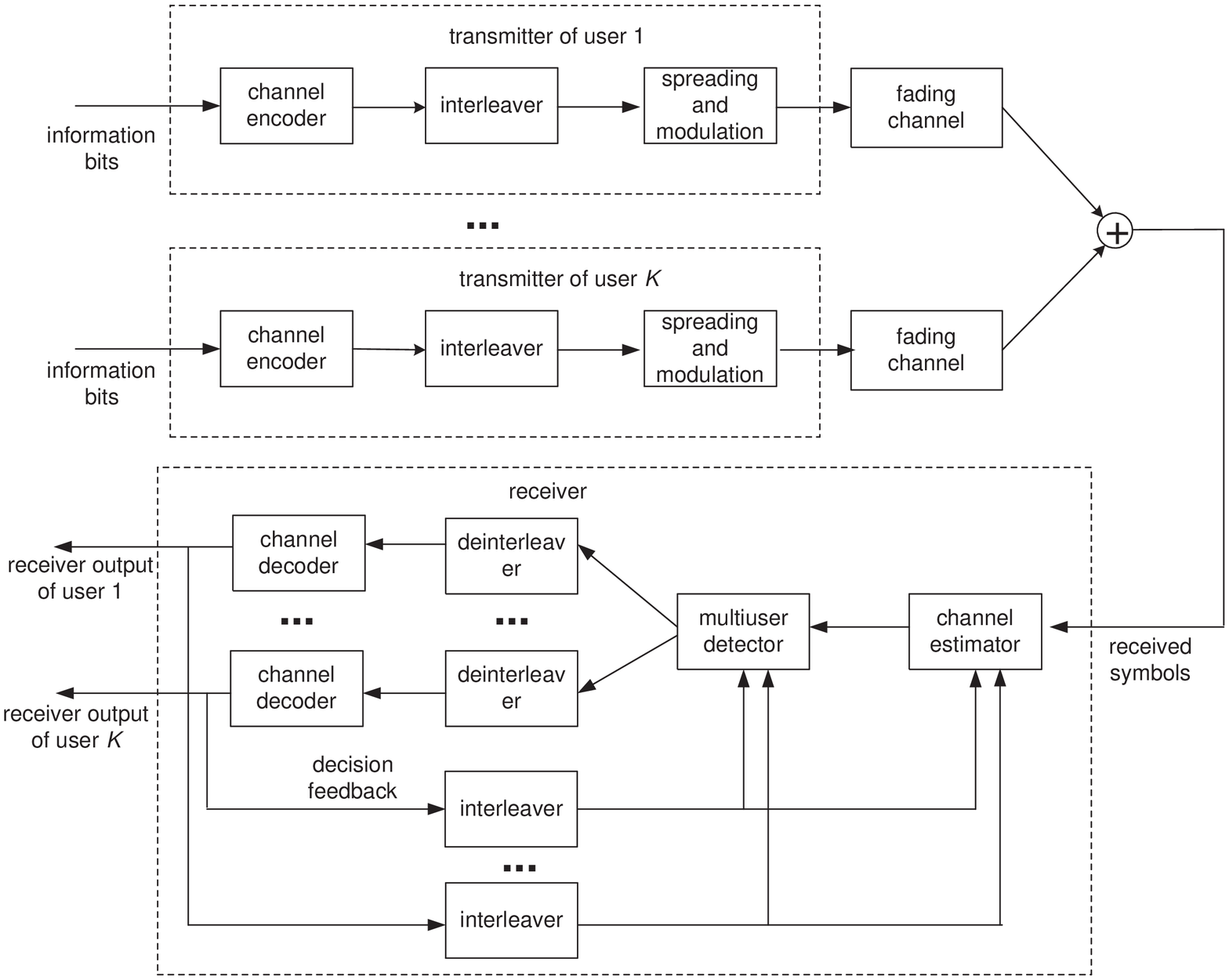}
  \caption{CDMA system with an iterative receiver}\label{}
\end{figure}

\begin{figure}
  \centering
  \includegraphics[scale=1]{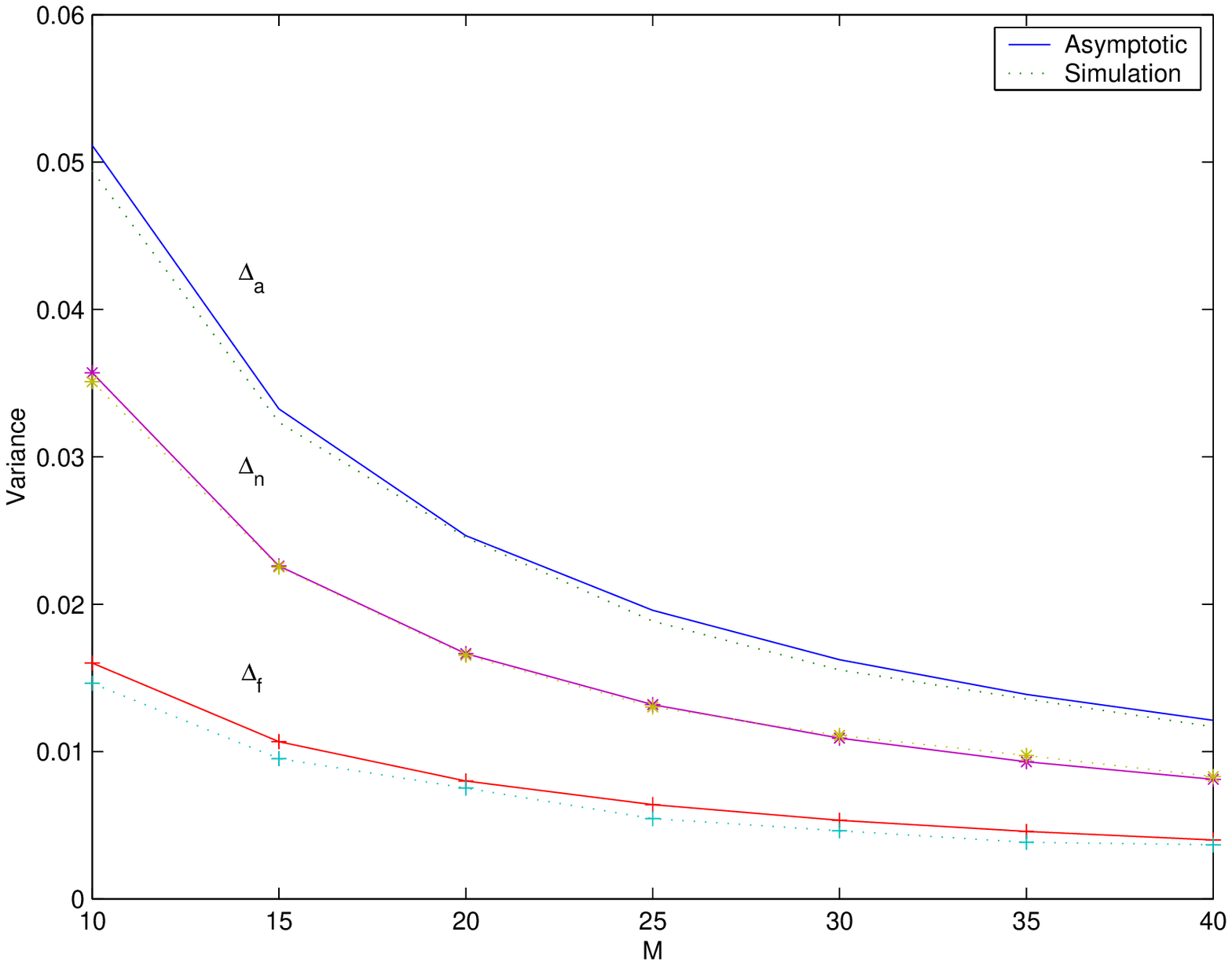}
  \caption{Average variance of channel estimates versus the coherence time $M$}\label{}
\end{figure}

\begin{figure}
  \centering
  \includegraphics[scale=1]{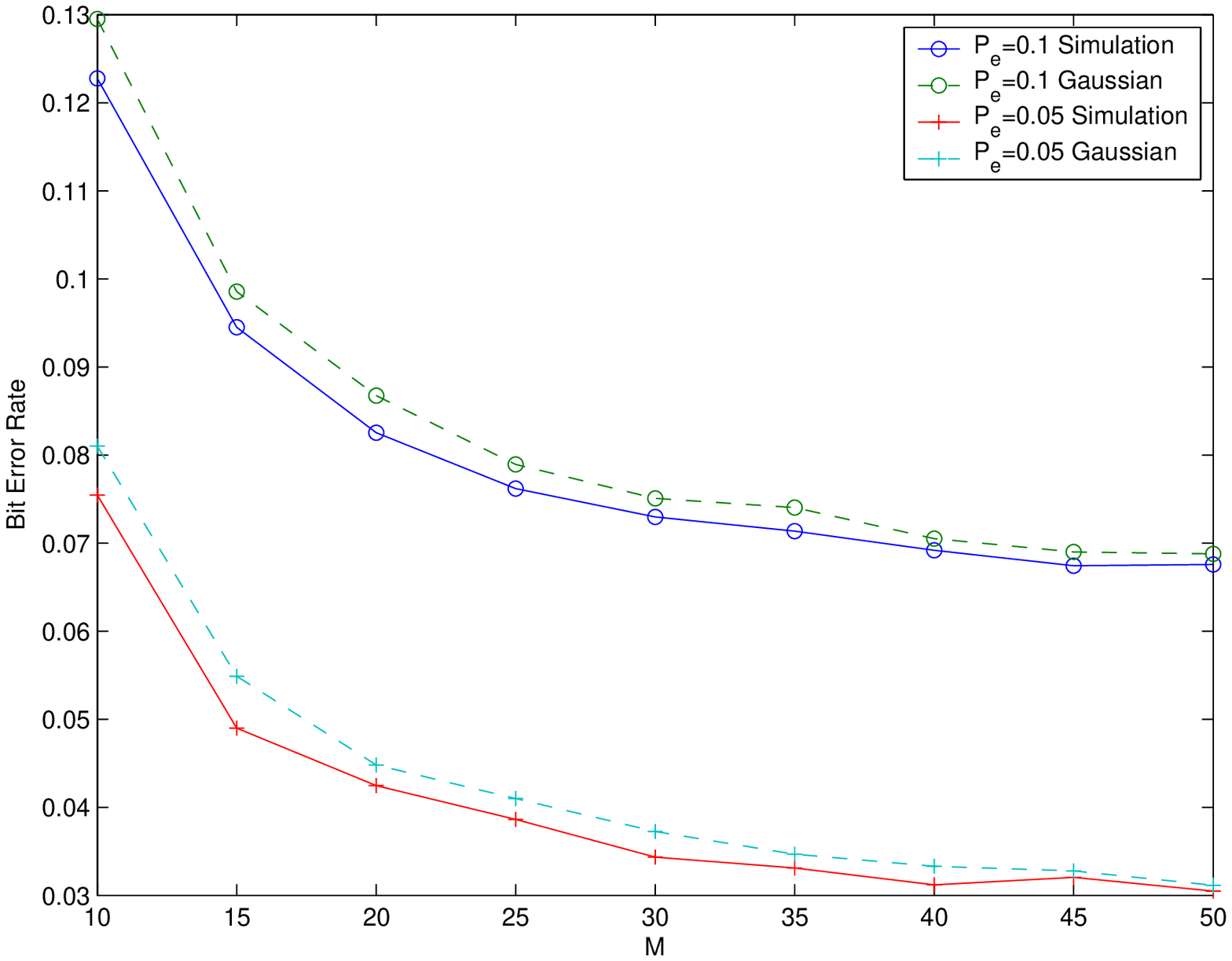}
  \caption{Comparison of simulated bit error rates and those obtained using a Gaussian approximation}\label{}
\end{figure}

\begin{figure}
  \centering
  \includegraphics[scale=1]{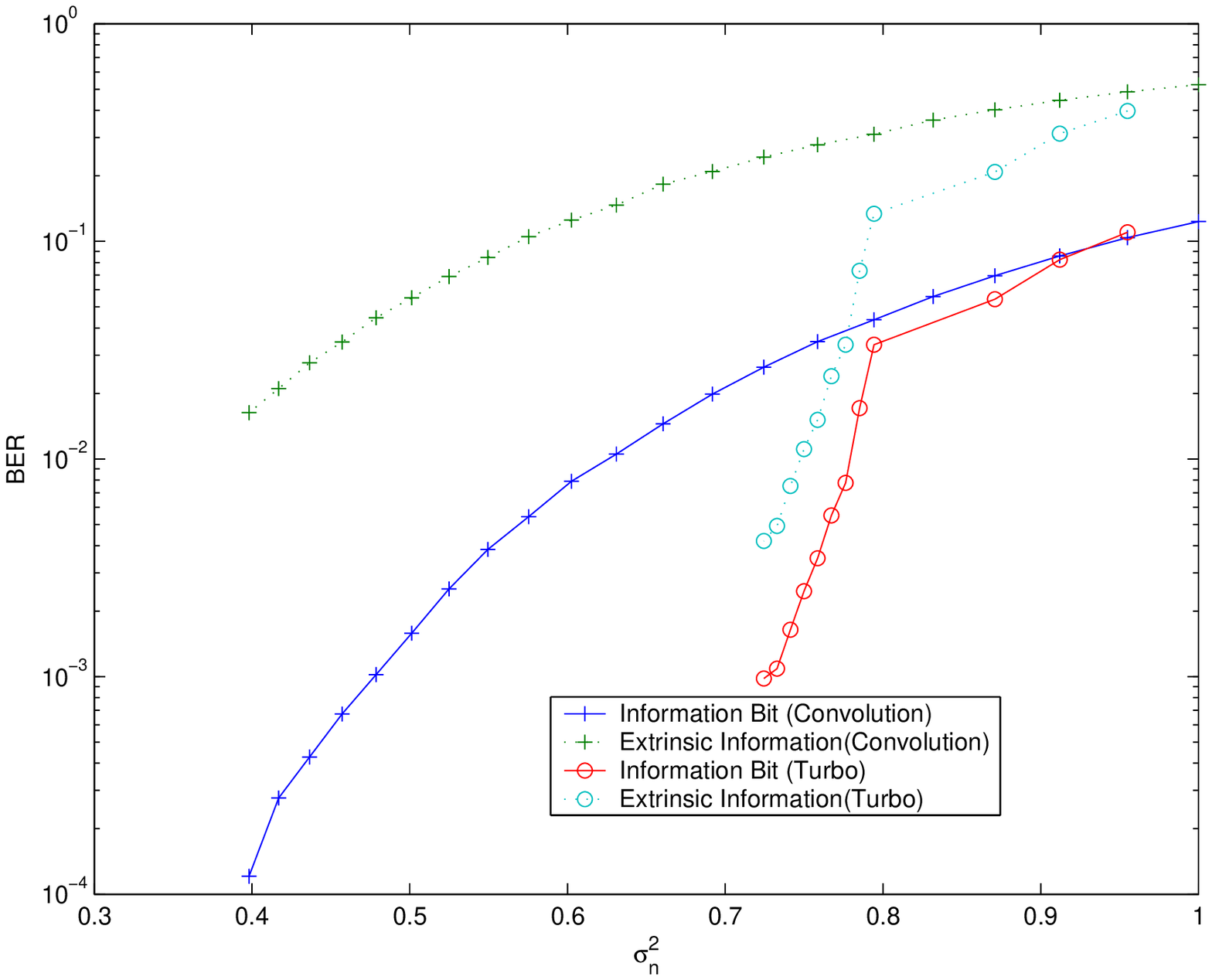}
  \caption{Performance of channel codes used in the numerical results, where the input SNR = $\frac{1}{\sigma_n^2}$}\label{}
\end{figure}

\begin{figure}
  \centering
  \includegraphics[scale=1]{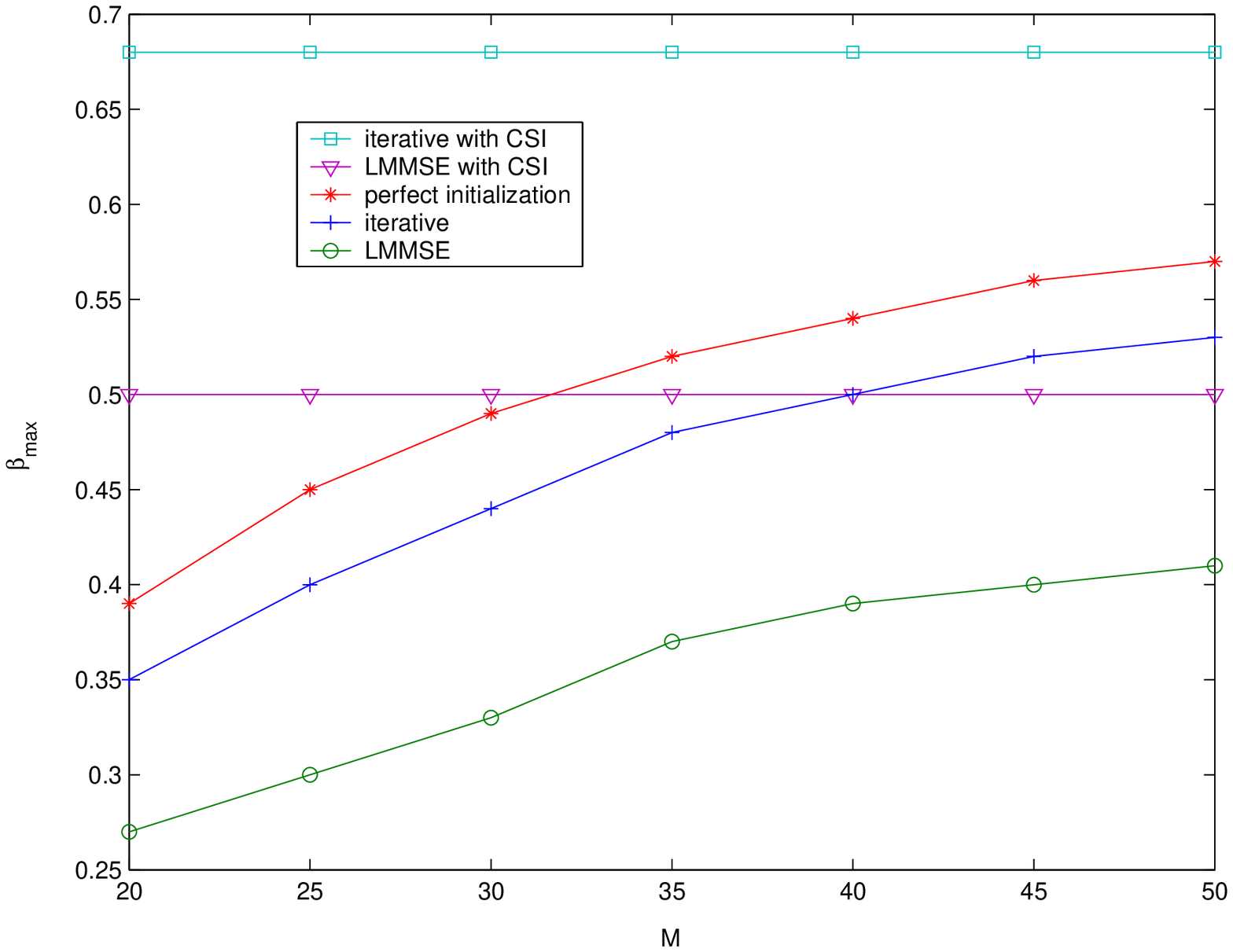}
  \caption{Maximum load of systems with convolutional codes}\label{}
\end{figure}

\begin{figure}
  \centering
  \includegraphics[scale=1]{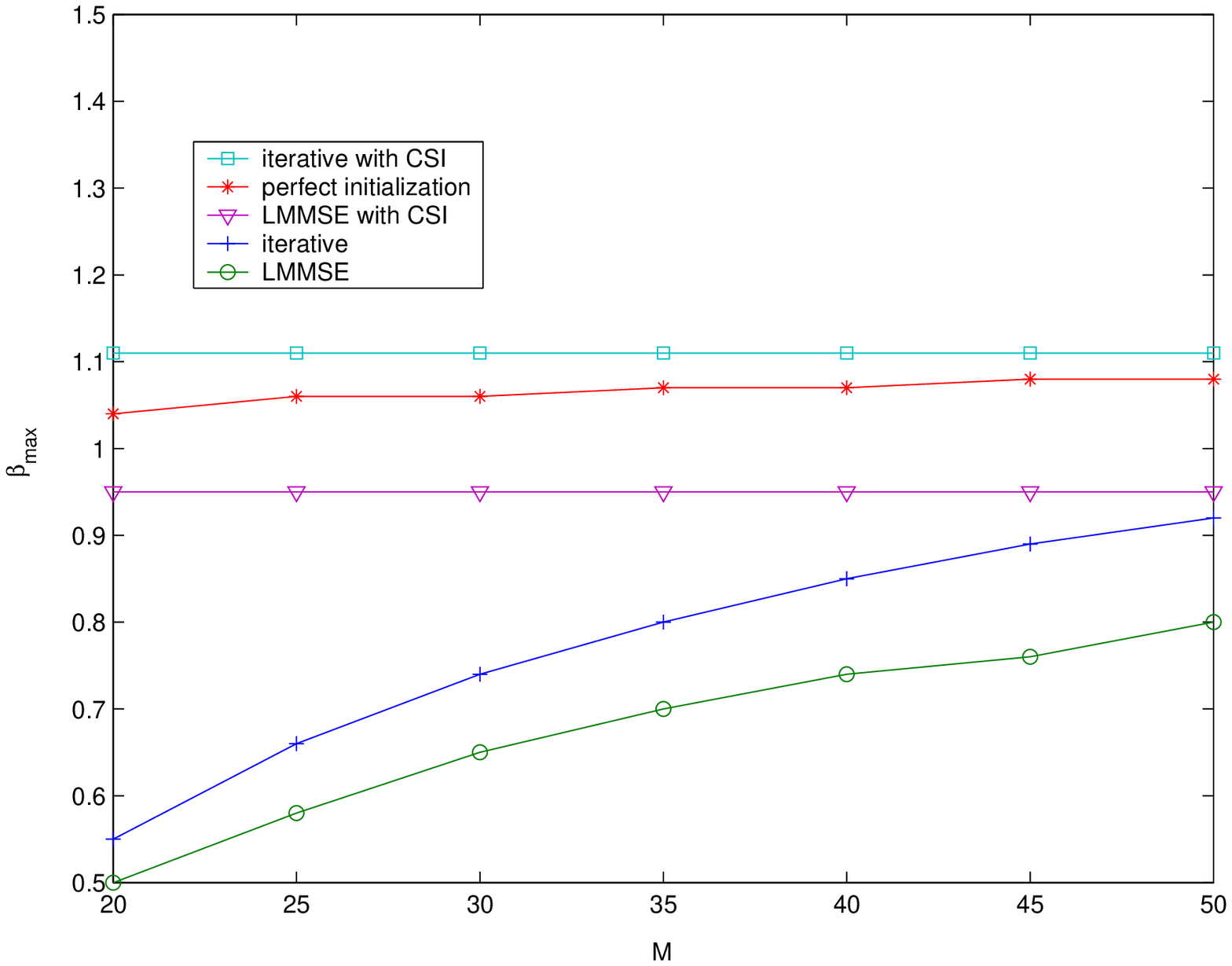}
  \caption{Maximum load of systems with turbo codes}\label{}
\end{figure}


\begin{thebibliography}{1}
\bibitem{Alex2000}
P. Alexander, A. Grant and M. C. Reed, ``Iterative detection of
code-division multiple-acess with error control coding,'' {\em
European Trans. Telecommun.}, Vol.~9,
  pp.~419--426, Aug. 1998.

\bibitem{AlexISSS2000}
P. Alexander and A. Grant, ``Iterative channel and information sequence estimation in CDMA,'' {\em Proceedings of
IEEE Sixth International Symposium on Spread Spectrum Techniques and Applications},
pp. 593--597, Parsippany, NJ, Sept. 2000.

\bibitem{Bai1993}
Z. D. Bai, J. W. Silverstein and Y. Q. Yin, ``A note on the largest
eigenvalue of a large dimensional sample covariance matrix ,'' {\em
Journal Multivariate Anal.}, Vol.~26,
  pp.~166--168, 1998.


\bibitem{BCJR}
L.~R.~Bahl, J.~Cocke, F.~Jelinek and J.~Raviv, ``Optimal decoding
of linear codes for minimizing symbol error rate,'' {\em IEEE
Trans. Inform. Theory}, Vol.~20,
  pp.~284--287, Aug. 1974.

\bibitem{Billingsley}
P. Billingsley, {\em Probability and Measure}.
\newblock John Wiley and Sons Inc, New York, US, 1995.

\bibitem{Buzzi2001}
S. Buzzi and H. V. Poor, ``Channel estimation and multiuser
detection in long-code DS/CDMA systems,'' {\em IEEE J. Select.
Areas Commun.}, Vol.~19,
  pp.~1476--1487, Aug. 2001.

\bibitem{Buzzi2003}
S. Buzzi, M. Lops and S. Sardellitti, ``Performance of iterative
data detection and channel estimation for single-antenna and
multiple-antenna wireless communications,'' {\em Proceedings of
2003 Asilomar Conference on Signals, Systems, and Computers},
Pacific Grove, CA, 2003.

\bibitem{Buzzi2004}
S. Buzzi, and H. V. Poor, ``A multi-pass approach to joint data and
channel estimation in long-code CDMA systems,'' {\em IEEE Trans.
Wireless Commun.}, Vol.~3,
  pp.~612--626, March, 2004.

\bibitem{Evans2000}
J. Evans and D. N. C. Tse, ``Large system performance of linear
multiuser receivers in multipath fading channels,'' {\em IEEE
Trans. Inform. Theory}, Vol.~46,
  pp.~2059--2078, Aug. 2000.

\bibitem{Golub}
G. H. Golub and C. F. Van Loan, {\em Matrix Computations}.
\newblock The Johns Hopkins University Press, Baltimore, MD, 1983.

\bibitem{Horm}
R. A. Horn and C. R. Johnson, {\em Matrix Analysis}.
\newblock Cambridge University Press, Cambridge, UK, 1985.

\bibitem{Hou2006}
J. Hou, J. E. Smee, H. D. Pfister and S. Tomasin, ``Implementing
interference cancellation to increase the EV-DO Rev A reverse link
capacity,'' {\em IEEE Commun. Mag.}, Vol.~44,
  pp.~96--102, Feb. 2006.


\bibitem{Kobayashi2001}
M. Kobayashi, J. Boutros and G. Caire, ``Successive interference
cancellation with SISO decoding and EM channel estimation,'' {\em
IEEE J. Select. Areas Commun.}, Vol.~19,
  pp.~1450--1460, Aug. 2001.

\bibitem{Kreyszig1998} E. Kreyszig, {\em Introductory Functional Analysis With Applications}.
\newblock John Wiley and Sons Inc, New York, 1989.

\bibitem{Lampe2002}
A. Lampe, ``Iterative multiuser detection with integrated channel estimation for coded DS-CDMA,'' {\em IEEE
Trans. Commun.}, Vol.~50, no. 8,
  pp.~1217--1223, Aug. 2002.


\bibitem{Laot2001}
C. Laot, A. Glavieux and J. Labat, ``Turbo equalization: Adaptive
equalization and channel decoding jointly optimized,'' {\em IEEE
J. Select. Areas Commun.}, Vol.~19,
  pp.~1744--1752, Sept. 2001.

\bibitem{LiHuCISS2003}
H.~Li and H. V. Poor, ``Performance of channel estimation in long
code DS-CDMA with and without decision feedback,'' {\em
Proceedings of the 2003 Conference on Information Sciences and
Systems}, The Johns Hopkins University, Baltimore, MD, March 2003.

\bibitem{LiHuEuroSip2005}
H.~Li and H. V. Poor, ``Impact of channel estimation error on
multiuser detection via the replica method,'' {\em EURASIP Journal
on Wireless Communicaitons and Networking}, Vol.2005,
pp.~175--186, May 2005.

\bibitem{Linbo2001}
L.~Li, A. M. Tulino and S. Verd\'{u}, ``Asymptotic eigenvalue
moments for linear multiuser detection,'' {\em Communications in
Information and Systems}, Vol.1, no. 3, pp. 273 -- 304, Sept.
2001.

\bibitem{Loncar2004}
M. Loncar, R. M\"{u}ller, J. Wehinger, C. Mecklenbraeuker, and T. Abe, ``Iterative channel estimation and data detection in
frequency-selective fading MIMO channels,'' {\em European Transactions on Telecommunications}, Vol.15, no.53, pp.459--470, Sept./Oct.
2004.

\bibitem{Lupas1989}
R.~Lupas and S.~Verd\'{u}, ``Linear multiuser detectors for
synthronous code-division multiple-acess channels,'' {\em IEEE
Trans. Inform. Theory}, Vol.~35,
  pp.~123--136, Aug. 1989.

\bibitem{Moher}
M. Moher, ``An iterative multiuser decoder for near capacity communications,'' {\em IEEE
Trans. Commun.}, Vol.~46, no.7,
  pp.~870--880, July 1998.


\bibitem{Niu2003}
H. Niu and J. A. Ritcey, ``Iterative channel estimation and
decoding of pilot symbol assisted LDPC coded QAM over flat fading
channels,'' {\em Proceedings of 2003 Asilomar Conference on
Signals, Systems, and Computers}, Pacific Grove, CA, 2003,

\bibitem{Poor2004}
H. V. Poor, ``Iterative multiuser detection,'' {\em IEEE Signal Processing Magazine}, Vol.~21,
No,1,  pp.~81--86, 2004.


\bibitem{Raheli1995}
R. Raheli, A. Polydoros, and C. Tzou, ``Per-survivor processing: A
general approach to MLSE in uncertain environments,'' {\em IEEE
Trans. Commun.}, Vol.~43,
  pp.~354--364, Feb./March/April, 1995.

\bibitem{Reed1998}
M. C. Reed, C. B. Schlegel, P. D. Alexander and J. A. Asenstorfer, ``Iterative multiuser detection for CDMA with FEC: Near
single user performance,'' {\em IEEE
Trans. Commun.}, Vol.~46, no.12,
  pp.~1693--1699, Dec. 1998.


\bibitem{Valenti1998}
M. C. Valenti and B. D. Woerner, ``Iterative multiuser detection for
convolutionally coded asynchronous DS-CDMA,'' in {\em IEEE Int.
Symp. Personal, Indoor, Mobile Radio Commun.}, Boston, Sept. 1998.

\bibitem{Verdu1986}
S. Verd\'{u}, ``Minimum probability of error for asynchronous
Gaussian multiple-access channels,'' {\em IEEE Trans. Inform.
Theory}, Vol.~32,
  pp.~85--96, Jan. 1986.

\bibitem{Verdu1998}
S. Verd\'{u}, {\em Multiuser Detection}.
\newblock Cambridge University Press, Cambridge, UK, 1998.

\bibitem{wang2004}
X. Wang and H. V. Poor, {\em Wireless Communication Systems: Advanced Techniques for Signal Reception}.
\newblock Prentice-Hall, Upper Saddle River, NJ, 2004.

\bibitem{wang1999}
X. Wang and H. V. Poor, ``Iterative (turbo) soft interference cancellation and decoding for coded CDMA,'' {\em
IEEE Trans. Commun.}, Vol.~47, no.7,
  pp.~1046--1061, July 1999.

\bibitem{wang2000}
X. Wang and H. V. Poor, ``Subspace methods for blind channel
estimation and multiuser detection   in CDMA systems,'' {\em
Wireless Networks}, Vol.~6,
  pp.~59--71, Feb. 2000.

\bibitem{Xu2000}
Z.~Xu and M.~K.Tsatsanis, ``Blind channel estimation for long code
multiuser CDMA systems,'' {\em IEEE Trans. Signal Processing},
Vol.~48,
  pp.~988--1001, Apr. 2000.
\end{thebibliography}
\end{document}

% END of IEEEtest.tex ************